\documentclass{aa}  
\usepackage{natbib}
\usepackage{graphicx}
\usepackage{multirow}
\usepackage{txfonts}
\newcommand*\chem[1]{\ensuremath{\mathrm{#1}}} 
\newcommand\Tstrut{\rule{0pt}{2.6ex}}         
\newcommand\Bstrut{\rule[-0.9ex]{0pt}{0pt}}   
\newcommand\Bstrutlong{\rule[-1.6ex]{0pt}{0pt}}

\begin{document} 

\title{The Near-Infrared Spectrograph (NIRSpec) on the James Webb Space Telescope}
\titlerunning{NIRSpec on the JWST}

\subtitle{IV. Capabilities and predicted performance for exoplanet characterization}

\author{S. M. Birkmann\inst{1}
\and P. Ferruit\inst{2}
\and G. Giardino\inst{3}
\and L. D. Nielsen\inst{4}
\and A. Garc\'ia Mu\~noz\inst{5}
\and S. Kendrew\inst{1}
\and B. J. Rauscher\inst{6}
\and T. L. Beck\inst{7}
\and C. Keyes\inst{7}
\and J. A. Valenti\inst{7}
\and P. Jakobsen\inst{8}
\and B. Dorner\inst{9}
\and C. Alves de Oliveira\inst{2}
\and S. Arribas\inst{10}
\and T. B\"oker\inst{1}
\and A. J. Bunker\inst{4}
\and S. Charlot\inst{11}
\and G. de Marchi\inst{12}
\and N. Kumari\inst{13}
\and M. L\'opez-Caniego\inst{14}
\and N. L\"utzgendorf\inst{1}
\and R. Maiolino\inst{15}
\and E. Manjavacas\inst{13}
\and A. Marston\inst{2}
\and S. H. Moseley\inst{16}
\and N. Prizkal\inst{13}
\and C. Proffitt\inst{7}
\and T. Rawle\inst{1}
\and H.-W. Rix\inst{9}
\and M. te Plate\inst{1}
\and E. Sabbi\inst{7}
\and M. Sirianni\inst{1}
\and C. J. Willott\inst{17}
\and P. Zeidler\inst{13}
}
\authorrunning{Birkmann et al.}

\institute{
European Space Agency, Space Telescope Science Institute, Baltimore, Maryland, USA \and
European Space Agency, European Space Astronomy Centre, Madrid, Spain \and
ATG Europe for the European Space Agency, European Space Research and Technology Centre, Noordwijk, The Netherlands \and
Department of Physics, University of Oxford, United Kingdom \and
CEA Saclay, Gif-sur-Yvette, France \and
NASA Goddard Space Flight Center, Observational Cosmology Laboratory, Greenbelt, USA \and
Space Telescope Science Institute, Baltimore, USA \and
Cosmic Dawn Center, Niels Bohr Institute, University of Copenhagen, Denmark \and
Airbus Defence and Space GmbH, Friedrichshafen, Germany. \and
Centro de Astrobiologia (CSIC-INTA), Departamento de Astrofisica, Madrid, Spain \and
Sorbonne Universit\'{e}, CNRS, UMR 7095, Institut d’Astrophysique de Paris, France \and
European, Space Agency, European Space Research and Technology Centre, Noordwijk, The Netherlands \and
AURA for the European Space Agency, Space Telescope Science Institute, Baltimore, Maryland, USA \and
Aurora Technology for the European Space Agency, European Space Astronomy Centre, Madrid, Spain \and
Kavli Institute for Cosmology, University of Cambridge, United Kingdom \and
Quantum Circuits, Inc., New Haven, Connecticut, USA \and
NRC Herzberg, Victoria, British Columbia, Canada
}

\date{Received November 05, 2021; accepted January 26, 2022}

 
  \abstract{The Near-Inrared Spectrograph (NIRSpec) on the James Webb Space Telescope (JWST) is a very versatile instrument, offering multiobject and integral field spectroscopy with varying spectral resolution ($\sim$30 to $\sim$3000) over a wide wavelength range from 0.6 to 5.3 micron, enabling scientists to study many science themes ranging from the first galaxies to bodies in our own Solar System. In addition to its integral field unit and support for multiobject spectroscopy, NIRSpec features several fixed slits and a wide aperture specifically designed to enable high precision time-series and transit as well as eclipse observations of exoplanets. In this paper we present its capabilities regarding time-series observations, in general, and transit and eclipse spectroscopy of exoplanets in particular. Due to JWST's large collecting area and NIRSpec's excellent throughput, spectral coverage, and detector performance, this mode will allow scientists to characterize the atmosphere of exoplanets with unprecedented sensitivity.}

   \maketitle
%

\section{Introduction}

The study of exoplanets has, during the last decades, become a key science research theme of the James Webb Space Telescope (JWST). This is a highly dynamic and fast-evolving field of research, driven by a wealth of new data from existing facilities, in particular from dedicated photometric surveys and telescopes, including, for example, WASP \citep[][]{Pollacco2006}, HAT \citep[][]{Bakos2002}, MEarth \citep[][]{Berta2012}, TRAPPIST, and its spin-off survey SPECULOOS \citep[][]{Sebastian2021} on the ground; CoRoT \citep[][]{Auvergne2009}, Kepler \citep[][]{Borucki2010}, and TESS \citep[][]{Ricker2014} in space, as well as from the Hubble and Spitzer Space Telescopes; among others. As of 2021, the NASA Exoplanet Archive\footnote{\url{https://exoplanetarchive.ipac.caltech.edu/}} contains over 4,500 confirmed exoplanets, most of which were detected through the transit method~\citep{Greene2019}. 

While huge progress has been made in generating and interpreting high-precision photometric time-series data, a full physical understanding of exoplanet formation, evolution, dynamics, and chemistry requires spectroscopic characterization. Exoplanet transit and eclipse spectroscopy is still in its infancy due to the highly challenging nature of the observations - both in terms of the technology capability as well as the interpretation of the data \citep[see, for example,][]{Pont2006,Sing2016}. Such spectroscopic characterization of exoplanetary atmospheres has only been performed for a few dozen planets as of yet~\citep{Greene2019}, see, for example, \citet{Bean2011}, \citet{Stevenson2014}, \citet{Kreidberg2018}, amongst others. A recent review by~\citet{Madhusudhan2019} gives a good overview of the current state-of-the-art and open questions in exoplanet atmosphere characterization using both photometric and spectroscopic techniques.

With its wide range of spectroscopic capabilities and superior sensitivity in the near- and mid-infrared, the JWST is expected to bring a new era in exoplanet atmosphere characterization~\citep{Greene2019}. Its operation at L2 allows for efficient, uninterrupted time-series observations of transiting exoplanets. JWST's wavelength range contains important signatures of numerous molecular species that are diagnostic of atmospheric composition and chemistry over a wide range of temperatures (e.g., H$_{2}$O, NH$_{3}$, CO, CO$_2$, CH$_4$). Long-duration spectroscopic phase curve observations will allow us to measure the emergent spectrum of the planet as a function of its orbital phase, thereby constraining the thermal and dynamical conditions in the atmosphere~\citep{Knutson2007,Stevenson2014, Madhusudhan2019}. 

Numerous investigations into JWST's capabilities for exoplanet atmosphere characterization have been presented in the literature. These have examined candidate selection criteria~\citep{Molliere2017, Louie2018, Kempton2018, Fortenbach2020}; performed trade-off studies between instrument modes~\citep{Beichman2014, Barstow2015, Greene2016, Batalha2017}; studied the impact of instrumental (and astrophysical) systematics on atmospheric retrieval results~\citep{Barstow2015, Taylor2020, Komacek2020}; and predicted the feasibility and outcomes of JWST observations for particular systems, such as Proxima Cen b~\citep{Kreidberg2016}, the TRAPPIST-1 system~\citep{Barstow2016, Batalha2018}, 55 Cancri e~\citep{Zilinskas2020}, WASP-63b~\citep{Kilpatrick2018}, WASP-43b~\citep{Venot2020}, amongst many others. The above studies collectively suggest that while spectroscopic characterization of true ``Earth-like'' planets will remain extremely challenging, JWST will deliver high quality transmission and emission spectra down to super-Earth masses. With the lifetime of JWST being more limited than that of the Hubble Space Telescope or typical ground-based facilities, such predictive investigations are extremely valuable for making optimal use of the observatory and its instrumentation.

Between the Guaranteed Time Observations (GTO) and Early Release Science (ERS) programs, over 800 hours of JWST cycle 1 observing time have already been allocated to observations of 27 unique transiting exoplanet systems~\citep{Greene2019}. The transiting exoplanet ERS program (PI: N. Batalha) is described in detail by~\citet{BeanERS}, with supplementary modeling results provided in~\citet{Venot2020}. Details of these targets and the proposed observations for both GTO and ERS programs are publicly available\footnote{\url{https://www.stsci.edu/jwst/science-execution/approved-programs}}. In addition to these, there are more than 30 approved cycle 1 general observer (GO) programs that facilitate transit, eclipse, and phase curve observations. In total, JWST will characterize the atmospheres of more than 50 exoplanets in its first year of operations, slightly more than even optimistic estimates \citep[e.g.,][]{Greene2019} predicted.

Many of these studies are enabled by community-led development of modeling and simulation tools, both for producing model spectra of transiting exoplanet systems and creating simulated JWST observations. Of note is the PandEXO package~\citep{pandexo}, which builds on the JWST exposure time calculator (ETC) \footnote{\url{https://jwst.etc.stsci.edu/}} code Pandeia~\citep{pontoppidan2016}.

The Near Infrared Spectrograph (NIRSpec) is one of four science instruments aboard the JWST, which was launched on 25 December 2021 on an Ariane 5 rocket. NIRSpec was developed by the European Space Agency (ESA) with Airbus Defence and Space (formerly EADS Astrium GmbH) as the prime contractor.

A detailed description of the NIRSpec instrument and its design is given by \citet[][Paper I henceforth]{Jakobsen2021}. In summary, NIRSpec provides spectroscopic capabilities in the $0.6$ to $5.3~\mathrm{\mu m}$ wavelength range with different spectral resolution and observing modes. NIRSpec has an integral field unit (IFU) for 3D-spectroscopy, a micro-shutter assembly (MSA) for multiobject spectroscopy of more than 100 sources simultaneously, and five slits (or apertures) for high-contrast and precision spectroscopy of individual sources. The instrument features seven dispersers (a double-pass prism and six gratings) that are available in all modes. Light is detected by two HAWAII-2RG 2k x 2k HgCdTe sensor chip arrays (SCAs) manufactured by Teledyne Scientific \& Imaging that form the focal plane assembly (FPA). The detectors are read nondestructively ``up-the-ramp'' \citep[see, for example,][]{Rauscher+2007} and will be operated at a temperature close to 42.8\,K in flight.

NIRSpec's multiobject spectroscopy (MOS) mode is presented by \citet[][Paper II]{Ferruit2021}. \citet[][Paper III]{Boeker2021} provides details on observing with the NIRSpec IFU. In this paper, we focus on NIRSpec's capabilities in characterizing exoplanets by means of transit or occultation spectroscopy and phase curve observations. The instrument features a dedicated observation mode that enables these bright object time-series observations as described in Section~\ref{sec:cap}. Section~\ref{sec:perf} focuses on brightness limits and performance estimates. Sources of potential systematic errors are described in Section~\ref{sec:sys}, and a few case studies are presented in Section~\ref{sec:examples}. The peculiarities of planning NIRSpec exoplanet transit and occultation observations are discussed in Section~\ref{sec:planning}. In Section~\ref{sec:pipe} we briefly describe the data pipeline for times-series observations, and give a short overall summary in Section~\ref{sec:con}.


\section{Capabilities for exoplanet characterization}
\label{sec:cap}

\subsection{Bright object time-series (BOTS) observations}

NIRSpec features a $1.6"\times 1.6"$ wide aperture (called S1600A1 hereafter) that was specifically introduced to enable high precision time-series observations of bright targets. The aperture was designed to be wide enough to limit aperture losses and the associated small variations in throughput due to jitter and drift of the target. On the other hand, it is small enough so that sky background and possible contamination by other nearby sources in the field of view usually are not an issue, as is often the case for slitless spectrographs. In total, NIRSpec has five apertures of various widths and length, but only S1600A1 is currently supported for time-series observations with exposures longer than 10,000 seconds.

\subsection{Detector readout and subarrays} \label{sec:detector}

As for the other JWST near-infrared instruments and detectors, a NIRSpec exposure consists of $n_{int}$ integrations ($1\leq n_{int}\leq 65535$). Each integration has one or more frames (individual reads), which are then averaged into groups on board. Each group can have one or more frames. At the beginning of each integration, the pixels are reset individually at the same cadence at which they are read out. Resetting the detector therefore takes the same amount of time as reading a frame of data.

Because most host stars of transiting exoplanets that will be observed with JWST/NIRSpec are bright, the detectors are read out in subarray mode, that is only a fraction of the 2048 x 2048 array is actually read ``up-the-ramp'' during integrations. Fewer pixels to be read means shorter frame times, and therefore more reads can be obtained before a source of a given brightness saturates the detector. The subarray mode in NIRSpec features a gain of approximately 1.4 e$^-$/DN (compared to $\sim 1$ e$^-$/DN for full frame readout mode, where ``DN'' stands for ``data number'' measured in counts), in order to fully utilize the well depth of the detectors. Saturation limits and efficiency are discussed in more detail in Section~\ref{sec:perf}. It is expected that most programs will make use of the so called \verb+NRSRAPID+ read pattern, where each group consists of one frame ($n_f = 1$), that is no on-board frame averaging is used and thus frame times and groups times are identical ($t_f = t_g$).

The available subarrays and their frame times are listed in Table~\ref{tab:subarrays}. After an exposure is complete, the NIRSpec detectors will go back into full frame idle mode, resetting all detector pixels periodically. There is some thermal settling related to this change from subarray to full frame mode and therefore, ideally, BOTS observations will take one exposure with as many integrations as needed for the observation. The maximum exposure time with the maximum number of integrations will depend on the subarray and the number of groups ($n_g$) per integration. For the shortest integrations in \verb+NRSRAPID+ mode ($n_g = 1$, two frame times: reset - read) the maximum exposure time is also given in Table~\ref{tab:subarrays}. It is longer for integrations with more groups, but there is currently a limitation of the total number of frames in one exposure (196,608). Multiple exposures can be taken in case a single one is not long enough to cover the entire event.

\begin{table}
\caption{Available subarrays for bright object time-series (BOTS) observations.}        
\label{tab:subarrays}
\centering          
\begin{tabular}{c c r@{.}l r@{.}l}
\hline\hline       
Subarray &Size [pixel] & \multicolumn{2}{c}{$t_g$ [s]} & \multicolumn{2}{c}{$t_{exp}$ [h]} \Tstrut\Bstrut\\ 
\hline                    
   SUB2048 & 32$\times$2048 & 0 & 90156 & 32 & 8\Tstrut\\  
   SUB1024A/B & 32$\times$1024 & 0 & 45100 & 16 & 4\\
   SUB512 & 32$\times$512 & 0 & 22572 & 8 & 22\\
   SUB512S & 16$\times$512 & 0 & 14364 & 5 & 23\Bstrut\\
\hline                  
\end{tabular}
\tablefoot{\scriptsize The subarray size is given in number of pixels along spatial $\times$ spectral axes. $t_g$ denotes the read time for a single group/frame in NRSRAPID mode (no frame averaging). $t_{exp}$ is the maximum exposure time for 65,535 of the shortest possible integrations ($n_g = 1$, $t_{int}(min) = 2\times t_g$) and will be longer for integrations with $n_g > 1$. SUB1024A and SUB1024B are two distinct subarrays with different locations on the detectors, but with the same dimension and therefore identical frame times.}
\end{table}

\subsection{Wavelength coverage and spectral resolution}

All seven NIRSpec dispersers are available for BOTS observations. The prism offers full wavelength coverage from 0.6 through 5.3\,$\mu$m in a single setup, with a spectral resolution that varies from 30 to 300. The medium and high resolution gratings span the 0.7 - 5.2\,$\mu$m range in four setups, with a spectral resolution of about 1000 and 2700, respectively (see Paper I for more details). Due to the physical separation between the two NIRSpec detectors (designated NRS1 and NRS2), the spectra of the high resolution gratings have a gap close to the middle of their bands. Spectra with the prism and the medium resolution gratings fall entirely onto NRS1. Table~\ref{tab:modes} lists the currently supported disperser and filter combinations and their wavelength coverage. For the gratings, the latter also depends on the selected subarray. The \verb+SUB512(S)+ subarray gives complete wavelength coverage for the prism and the \verb+SUB2048+ subarray for the gratings, apart from the wavelength gap due to the spacing between the detectors. The \verb+SUB1024A/B+ subarrays provide shorter frame times (see Table~\ref{tab:subarrays}) for observations with the gratings, thus allowing brighter sources, at the expense of reduced wavelength coverage. \verb+SUB1024A+ covers the short wavelength end for NRS1 and the long wavelength end for NRS2, while \verb+1024B+ covers the long wavelength end on NRS1 and the short wavelength end on NRS2. For the high resolution gratings this means that \verb+SUB1024A+ covers the extreme ends of the \verb+SUB2048+ range, whereas \verb+SUB1024B+ covers the central part.

\begin{table}
\caption{Available filter and disperser combinations for time-series observations with the S1600A1 aperture}
\label{tab:modes}      
\centering          
\begin{tabular}{c r@{.}l r@{.}l r@{.}l r@{.}l r@{.}l r@{.}l}     
\hline\hline    
\multirow{2}{*}{Disperser/Filter} & \multicolumn{12}{c}{Wavelength coverage [$\mu$m]}\Tstrut\\
& \multicolumn{6}{c}{NRS1}&\multicolumn{6}{c}{NRS2}\Bstrut\\

\hline                
   PRISM/CLEAR & 0 & 60 & \multicolumn{2}{c}{} & 5 & 30\Tstrut\\  
   & \multicolumn{6}{c}{\raisebox{2ex}{\smash{$\underbrace{\makebox[7em]{}}_{\mathrm{SUB512\;\&\;SUB512S}}$}}} \\
   \\
   & \multicolumn{6}{c}{\raisebox{-1ex}{\smash{$\overbrace{\makebox[7em]{}}^{\mathrm{SUB2048}}$}}} & \multicolumn{6}{c}{\raisebox{-1ex}{\smash{$\overbrace{\makebox[7em]{}}^{\mathrm{\Large SUB2048}}$}}}\\
   G140M/F070LP & 0 & 70 & 1 & 22 & 1 & 27\\ 
   G140M/F100LP & 0 & 97 & 1 & 22 & 1 & 87\\ 
   G235M/F170LP & 1 & 66 & 2 & 03 & 3 & 12\\ 
   G395M/F290LP & 2 & 87 & 3 & 35 & 5 & 18\\ 
   G140H/F070LP & 0 & 82 & 1 & 07 & 1 & 27\\ 
   G140H/F100LP & 0 & 97 & 1 & 07 & 1 & 31 & 1 & 35 & 1 & 59 & 1 & 83\\ 
   G235H/F170LP & 1 & 66 & 1 & 79 & 2 & 20 & 2 & 27 & 2 & 67 & 3 & 07\\ 
   G395H/F290LP & 2 & 87 & 3 & 03 & 3 & 72 & 3 & 82 & 4 & 51 & 5 & 18\\ 
   & \multicolumn{12}{c}{\raisebox{2ex}{\smash{$\underbrace{\makebox[3.5em]{}}_{\mathrm{SUB1024A}} \underbrace{\makebox[3.5em]{}}_{\mathrm{SUB1024B}} \makebox[1.5em]{} \underbrace{\makebox[3.5em]{}}_{\mathrm{SUB1024B}} \underbrace{\makebox[3.5em]{}}_{\mathrm{SUB1024A}}$}}} \Bstrutlong\\
\hline
\end{tabular}
\tablefoot{\scriptsize Covered wavelength range for the uncontaminated first order. For the gratings, the SUB1024A and SUB1024B subarrays only partially cover the wavelength range that is spanned by the SUB2048 subarray. The exact position of the wavelength gap for the high resolution gratings will be re-measured in orbit, but also depends on the GWA tilt angle (see Paper I for details) and can thus vary slightly from observation to observation.}
\end{table}

\begin{figure}
\includegraphics[trim={1.0cm 0.3cm 1.3cm 1cm},clip,width=\hsize]{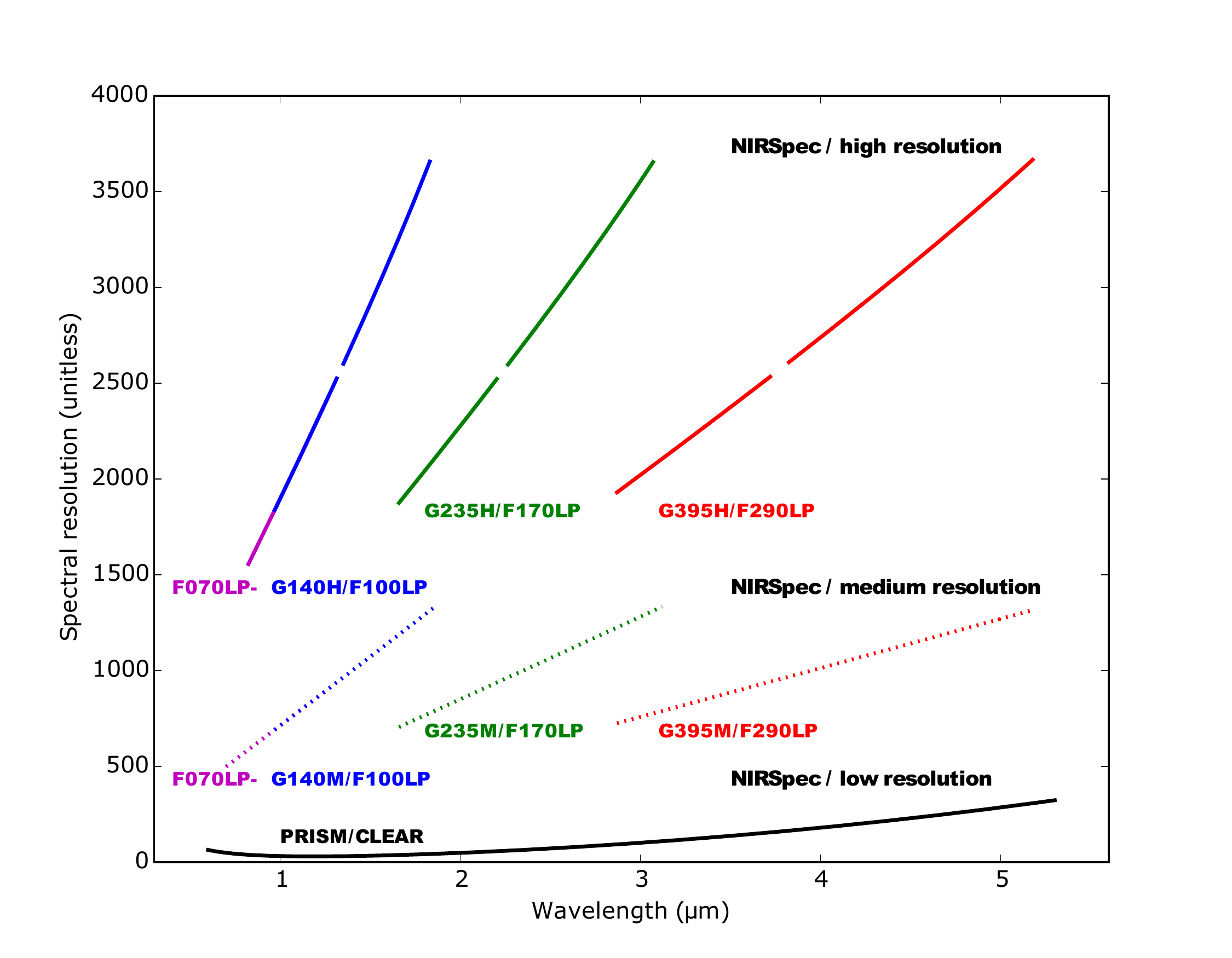}
\caption{\label{fig:resolution}Spectral resolution of the available NIRSpec dispersers. Wavelengths lost due to the small separation between the two detectors in the FPA are represented as a gap in the resolution curves for the high resolution gratings (for the S1600A1 aperture).}
\end{figure}

The wavelength-dependent spectral resolution of the different dispersers is shown in Figure~\ref{fig:resolution}. They refer to the nominal (i.e., uncontaminated from second order) ranges (see Paper I).

\subsection{BOTS detector data}

\begin{figure}
\includegraphics[width=\hsize]{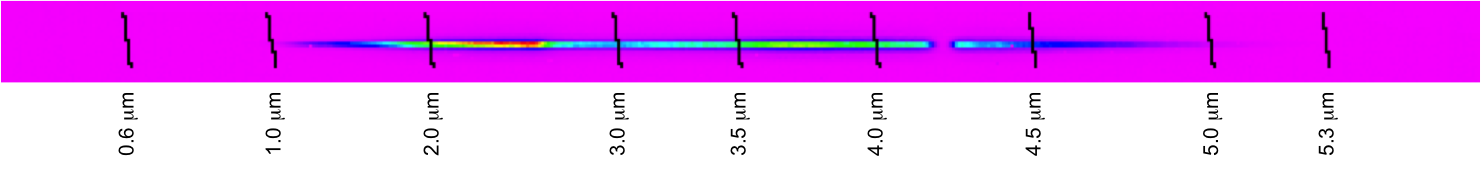}
\caption{\label{fig:data_prism}2D unrectified spectrum of a point source taken with PRISM/CLEAR and the SUB512 (32x512) subarray. The lack of flux at $\sim 4.2\,\mu$m is due to a spectral feature in the source.}
\end{figure}

\begin{figure*}
\includegraphics[width=\hsize]{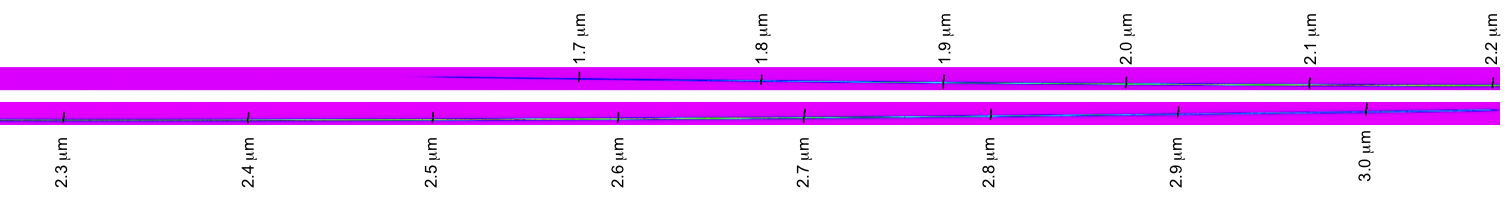}
\caption{\label{fig:data_grating}2D unrectified spectrum of a point source taken with F170LP/G235H and the SUB2048 (32x2048) subarray. The top panel shows the blue part of the spectrum recorded on the NRS1 detector, while the bottom panel shows the red part of the spectrum recorded simultaneously on the NRS2 detector. The slightly slanted (due to slit tilt, see Paper I for details) lines mark the wavelengths.}
\end{figure*}

Due to the NIRSpec spectrograph and camera optics, all NIRSpec spectra show a mild curvature and slit tilt. Both effects are discussed in more detail in Paper I. The curvature causes a change of the trace position inside the subarray as a function of wavelength and is the main reason why all subarrays have to be oversized with respect to the projected aperture size. For the prism this effect is very small, also due to the short trace, thus making the narrow \verb+SUB512S+ subarray possible. The slit tilt also affects all dispersers and results in nonperpendicular iso-wavelength lines with respect to the trace. Both effects are taken into account when computing and assigning the world coordinate system (WCS) to NIRSpec data by the processing pipeline (see Paper I), for example, the wavelength of every pixel and its relative position within the aperture.  
Figures~\ref{fig:data_prism} and \ref{fig:data_grating} show example spectra taken during ground testing with the prism and the G235H grating, respectively.

\subsection{Prospects for getting spectra of imaged exoplanets}

While this paper focuses on observing transiting exoplanets with NIRSpec, it must be noted that NIRSpec will also be capable of obtaining spectra of well separated (nontransiting) exoplanets that can be observed with direct imaging. This is achieved by placing the exoplanet into the IFU aperture, whereas the host star is placed outside the aperture and thus light from it will be highly suppressed, depending on the separation. An example for these kind of observations is an accepted GO program of the two exo-planets in the TYC 8998-760-1 multiplanetary system \citep[][]{Bohn2020}.

For systems with moderate contrast between planet(s) and host star, it will also be possible to have all objects inside the IFU field of view without significant detrimental effects, for example, stray light and charge diffusion in the detector. Examples include guaranteed time observations of, for instance, the TWA-27 \citep[][]{Chauvin2004} and HR8799 \citep[][]{Marois2010} systems scheduled for cycle 1. Results from ground testing indicate that NIRSpec has little internal stray light, and the limiting factor will be the flux in the wings of JWST's PSF that overlaps with the exoplanet signal. How much separation is needed will depend on the contrast between host star and planet, as well as the actual PSF quality and stray light performance of JWST/NIRSpec, which will be characterized in flight.

Because most direct imaging exoplanets are significantly fainter than their host stars, saturation is probably not that much of an issue for this science case, although it might impact the spectra of the host star if located within the IFU field of view. Sensitivity estimates for the NIRSpec IFU are given in Paper III and the saturation limits can be estimated using the JWST ETC.


\section{Performance} \label{sec:perf}

The NIRSpec Exoplanet Exposure Time Calculator \citep[NEETC,][]{Nielsen2016} has been developed by the instrument team to enable detailed investigations into the performance in terms of exoplanet observations. The NEETC is based on a radiometric model of NIRSpec, validated with throughput measurements obtained during  ground testing (Paper I), and is optimized to deal with BOTS. Only the noise floor consisting of shot noise (from the combined stellar signal and dark current) and readout noise are taken into account. The thermal noise when resetting the detector pixels (also referred to as kTC noise) is also considered when using the reset level to sample up-the-ramp, that is, for short ramps with very few reads. A detailed description of the noise model is given in \citet{Nielsen2016}. The NEETC was developed independently from PandEXO package~\citep{pandexo} and have been used to benchmark the NIRSpec-specific outputs of this widely used community tool.

The NEETC allows one to investigate saturation limits, wavelength coverage and signal-to-noise (S/N) per pixel while comparing the different observation configurations, for example, comparing medium and high resolution gratings rebinned to the same final resolution. The results presented in the following Sections, apart from Section~\ref{sec:sys} on sources of systematic errors, are based on NEETC calculations\footnote{More results and simulations can be found at \url{www.cosmos.esa.int/web/jwst-nirspec/exo-planets}}.

\subsection{Saturation and brightness limits} \label{sec:sat}
It is expected that exoplanets transiting bright host stars will be the preferred targets for JWST/NIRSpec, so the final S/N will be dominated by stellar photon noise. To predict when saturation will occur, we use realistic throughput curves (see Paper I) combined with a conservative (low) full-well value of $65,000~\mathrm{e^-}$ per pixel to ensure that we operate on the safe side of saturation. The shortest exposure possible (reset - read, as described in Section~\ref{sec:SNR}) is used along with the smallest subarray that still ensures full wavelength coverage (\verb=SUB512= for PRISM/CLEAR and \verb=SUB2048= for the gratings). 

\begin{figure}
\centering
\includegraphics[trim={0.0cm 0.0cm 1.0cm 1cm},clip,width=\hsize]{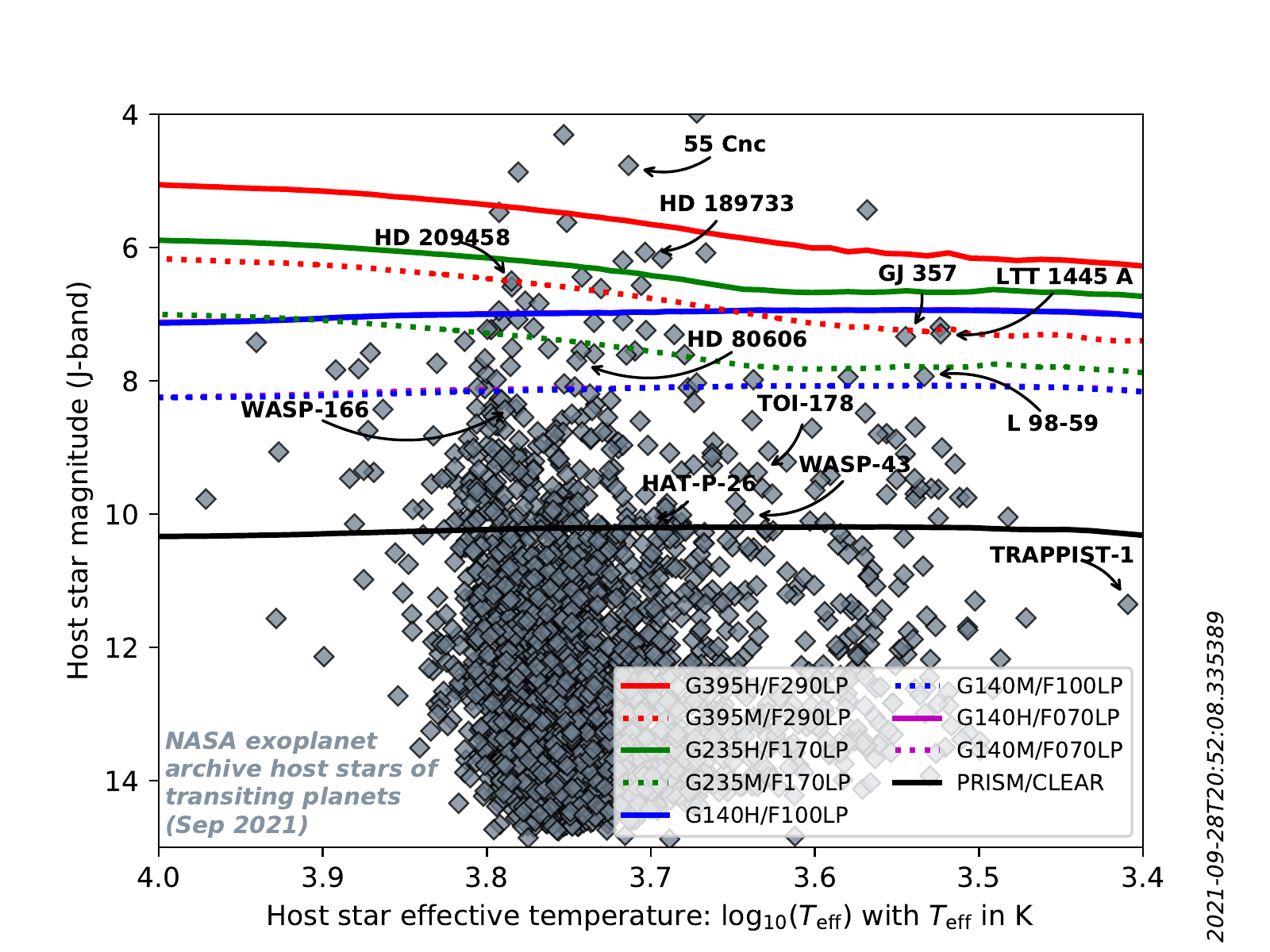}
\caption{Brightness limits for the available filter and disperser combinations in J-band Vega magnitudes. Calculated using the default subarrays, that is, SUB512 for PRISM/CLEAR and SUB2048 for all gratings. If a host star lies beneath the curve for a given configuration, that target can be observed without any saturation in the spectrum. It is still possible to observe host stars that fall above the curves, but with partial saturation in some wavelength ranges. An uncertainty of 20 \% should be considered when using these brightness limits.}
\label{fig:sat2}
\end{figure}

Figure \ref{fig:sat2} shows, for all 9 configurations, the limiting host star J-magnitude for which no saturation occurs anywhere in the spectrum as a function of host star temperature. PHOENIX stellar models \citep{Allard2011} with solar metallicity and surface gravity $\log(g)=4.5$ (cgs) have been used as host star templates to take the spectral energy distributions into account. We note that it is possible to observe brighter targets at the expense of wavelength coverage lost due to partial saturation in the spectrum. The curves have a 20\% uncertainty due to uncertainty in throughput. 

Host stars of known transiting exoplanets as at September 2021 are over-plotted in gray, with a few labeled by name. It is clear that most of these targets can be observed with JWST/NIRSpec even with the PRISM configuration and only a handful are out of reach for the high resolution gratings. For proposal preparation, observers are highly encouraged to use the official JWST exposure time calculator in order to assess number of groups before saturation / brightness limits for their targets.

\subsection{Signal-to-noise estimates} \label{sec:SNR}

\begin{figure}
\includegraphics[trim={1.3cm 0.3cm 1.3cm 0cm},clip,width=\hsize]{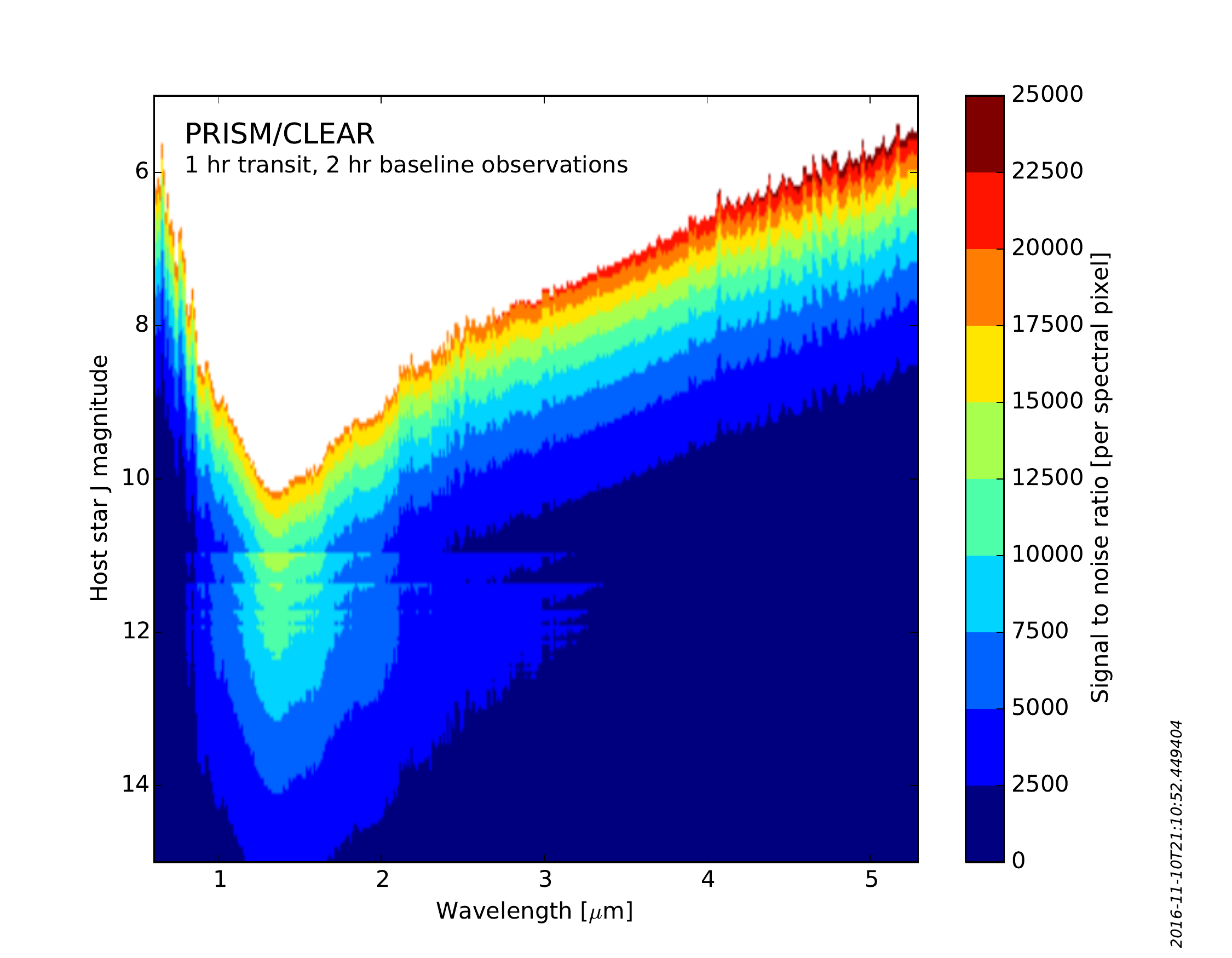}
\caption{Signal-to-noise estimates for the PRISM configuration based on a host star with effective temperature 3400~K and $\log(g)=4.5$ (cgs).}
\label{fig:snr_prism}
\end{figure}
To illustrate the typical S/N level for a single 1~hr transit (or occultation) combined with 2~hr out-of-eclipse baseline observations, we have run NEETC computations for a host star with $\log(g)=4.5$ (cgs), effective temperature $T_{\mathrm{eff}}=3400~\mathrm{K}$ and variable J magnitude. Only the noise floor containing photon and detector noise (i.e., no systematics) are taken into account and only results for the default subarrays are presented here. 

Figure \ref{fig:snr_prism} shows the S/N as a function of wavelength for such a transit or occultation observation using the PRISM/CLEAR configuration with subarray \verb=SUB512=. The white area represents saturation. A host star of magnitude $J=9$ will saturate from 1 to 2\,$\mathrm{\mu m}$ and have S/N ranging from 2,000 to 20,000, highly dependent on the wavelength.

The number of groups per integration is chosen to ensure the best possible duty cycle without saturating any parts of the detector. For targets so bright that some pixels will saturate in the first group, one group per integration is chosen (also referred to as read - reset). The discrete jumps in S/N seen as horizontal lines in Fig. \ref{fig:snr_prism} are due to changes in the number of groups per integration before saturation is met. For example, the sudden change in S/N at $J\sim 11$ indicates the regime where the duty cycle goes from 50 \% (each integration is made up of one reset and one read) to 66 \% (two groups for each reset). Again around magnitude 11.5 the duty cycle goes from 66 \% to 75\%. This effect is more pronounced close to the saturation limit and the steps therefore seem to disappear.

\begin{figure*}
\includegraphics[trim={1.3cm 0.2cm 1.3cm 0.5cm},clip,width=0.492\hsize]{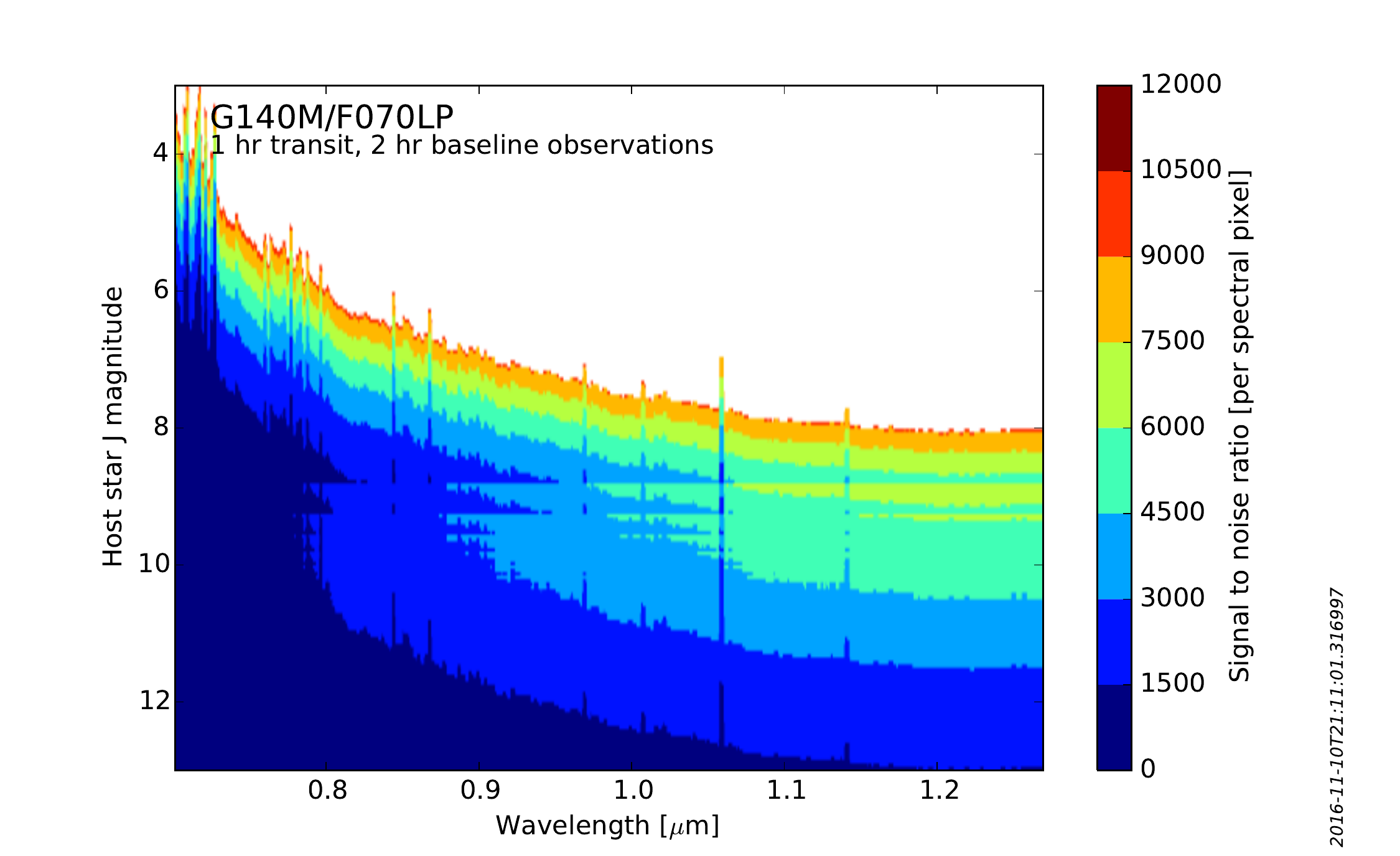}
\includegraphics[trim={1.3cm 0.2cm 1.3cm 0.5cm},clip,width=0.492\hsize]{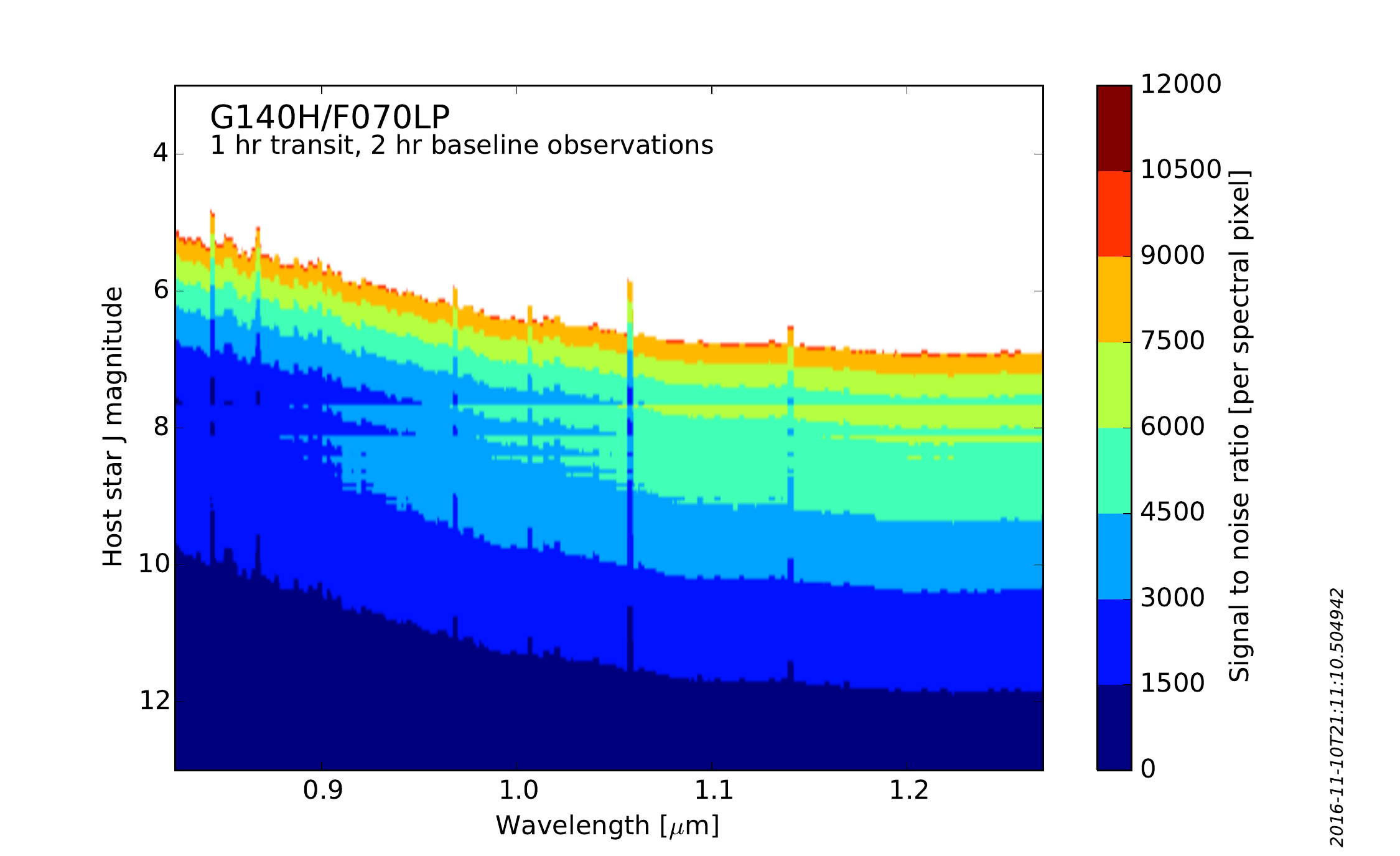}
\includegraphics[trim={1.3cm 0.2cm 1.3cm 0.5cm},clip,width=0.492\hsize]{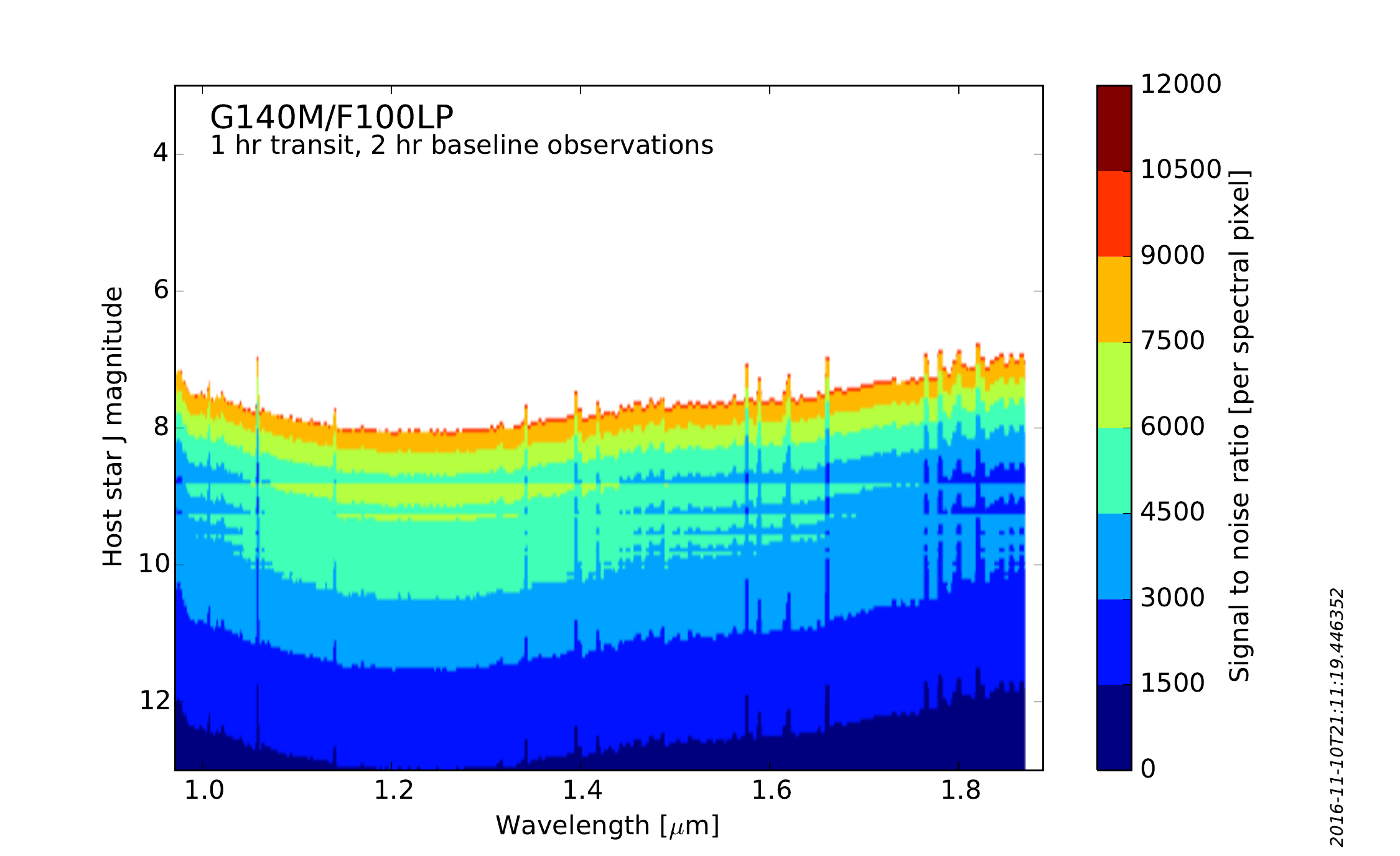}
\includegraphics[trim={1.3cm 0.2cm 1.3cm 0.5cm},clip,width=0.492\hsize]{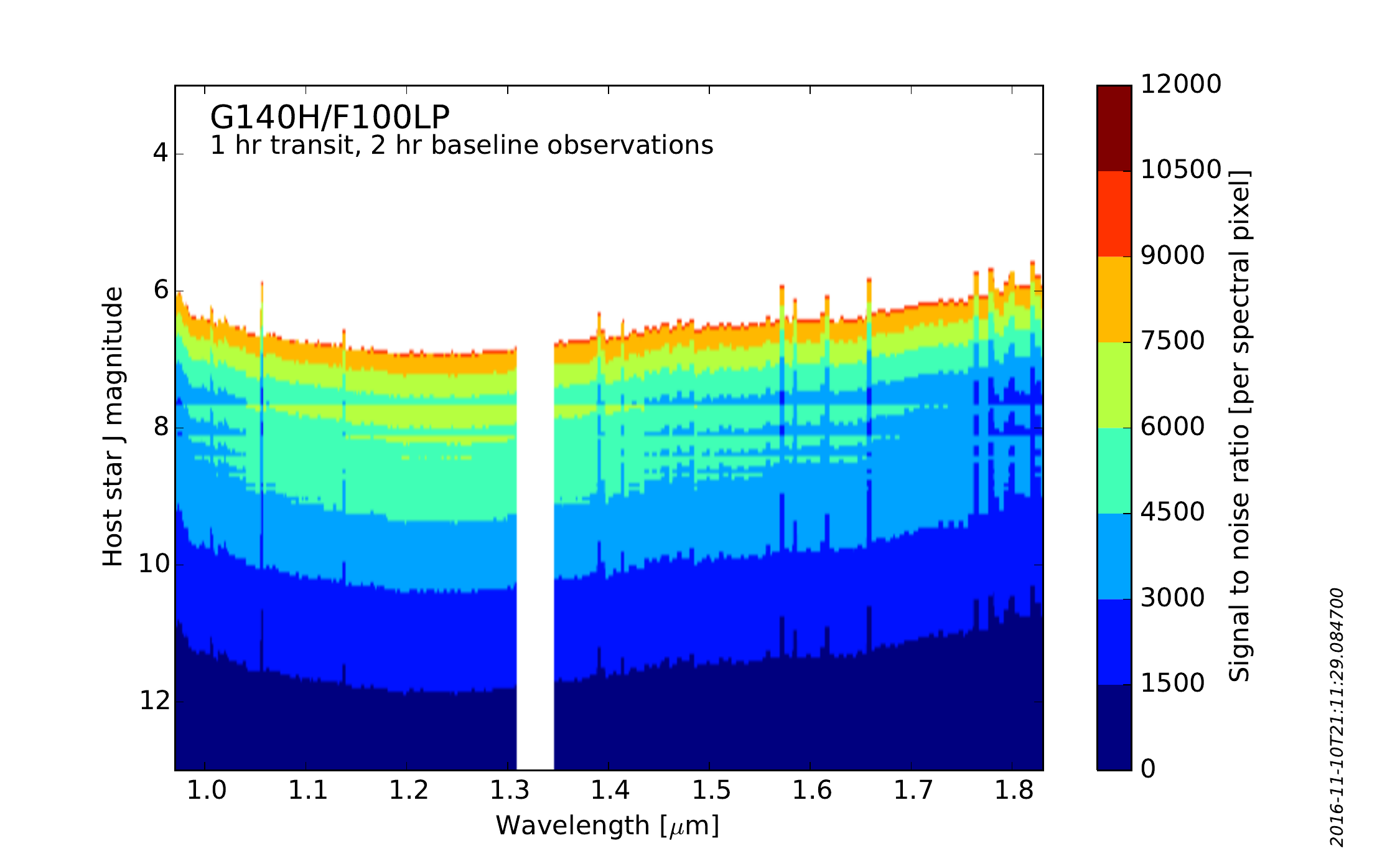}
\includegraphics[trim={1.3cm 0.2cm 1.3cm 0.5cm},clip,width=0.492\hsize]{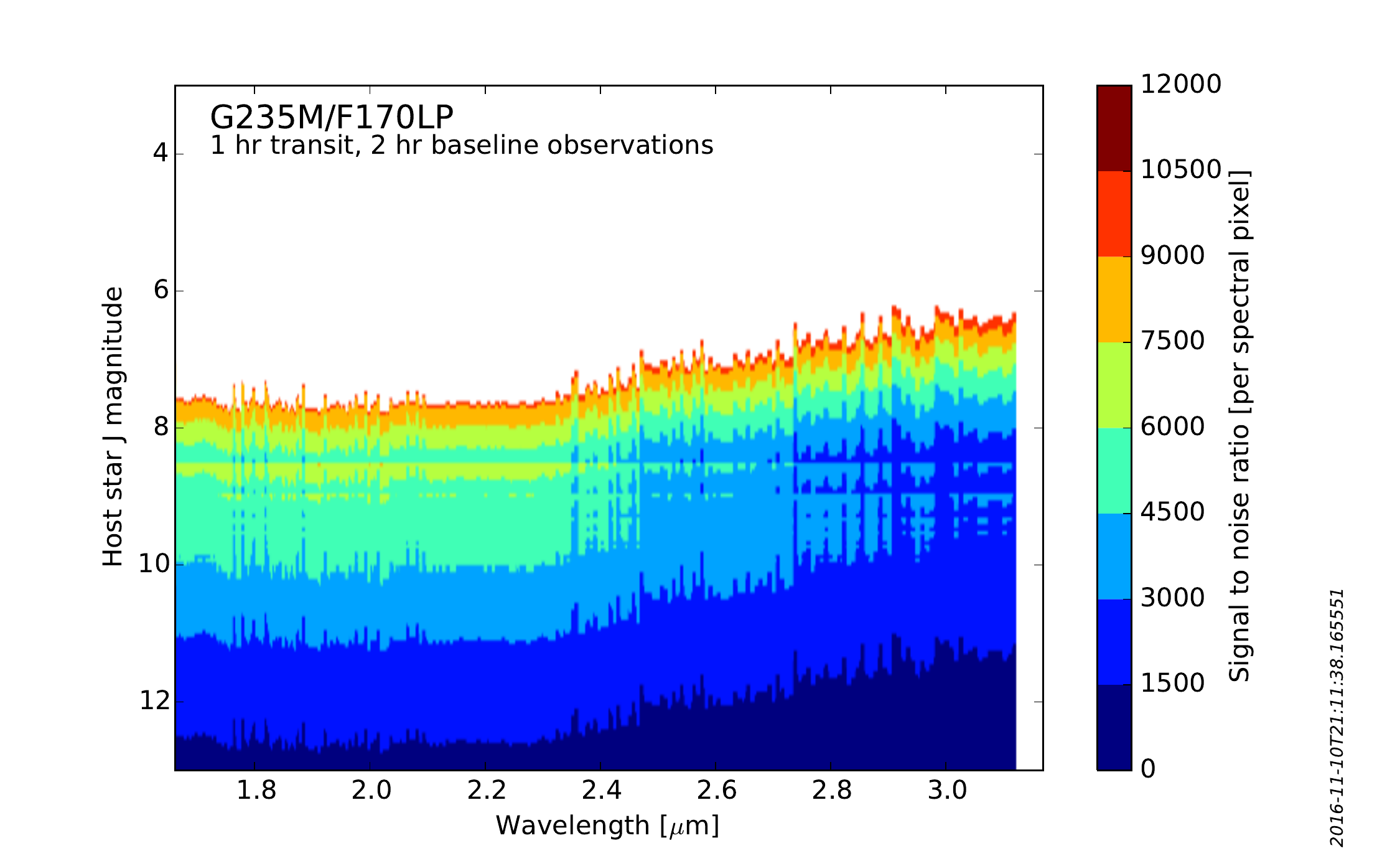}
\includegraphics[trim={1.3cm 0.2cm 1.3cm 0.5cm},clip,width=0.492\hsize]{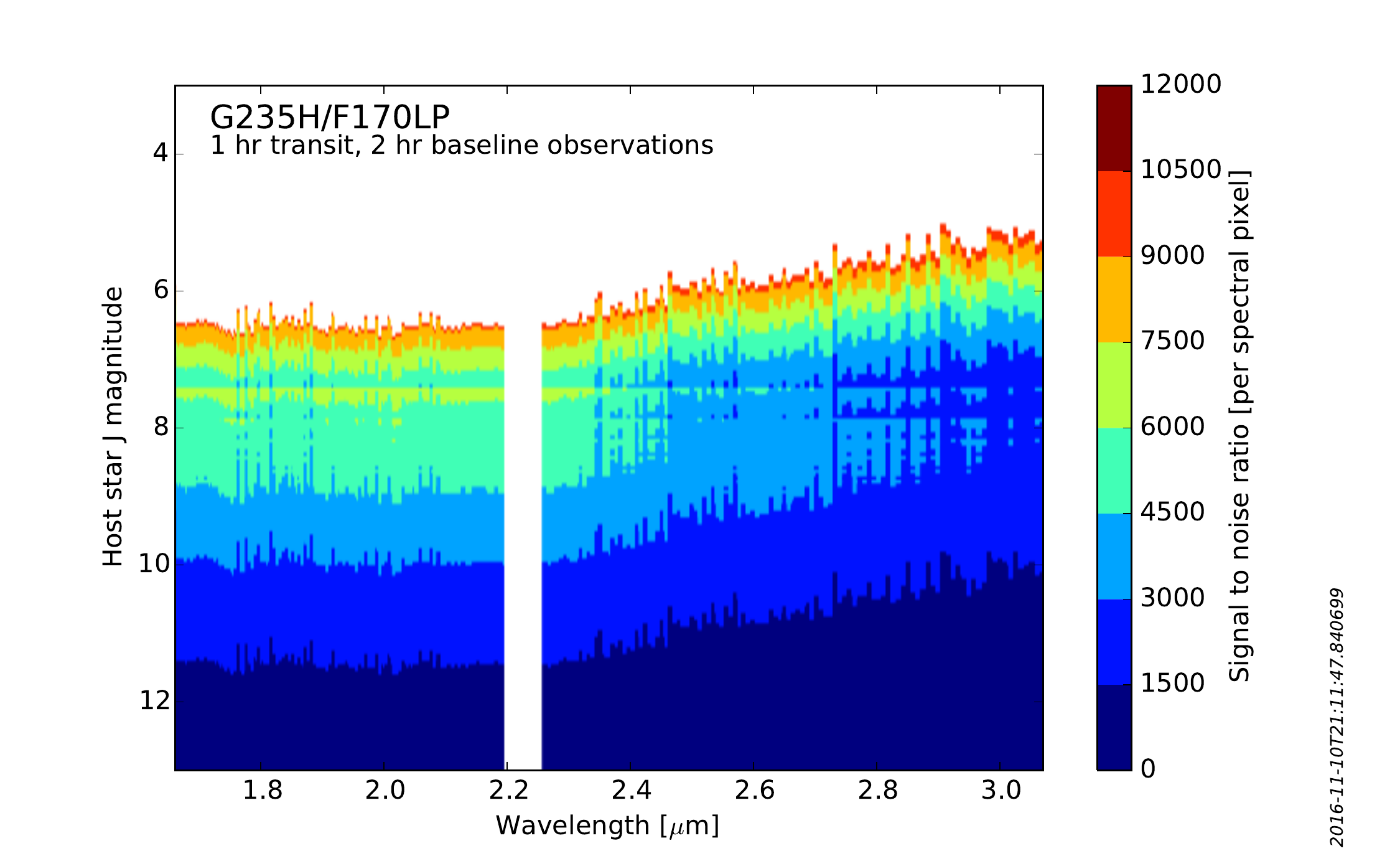}
\includegraphics[trim={1.3cm 0.2cm 1.3cm 0.5cm},clip,width=0.492\hsize]{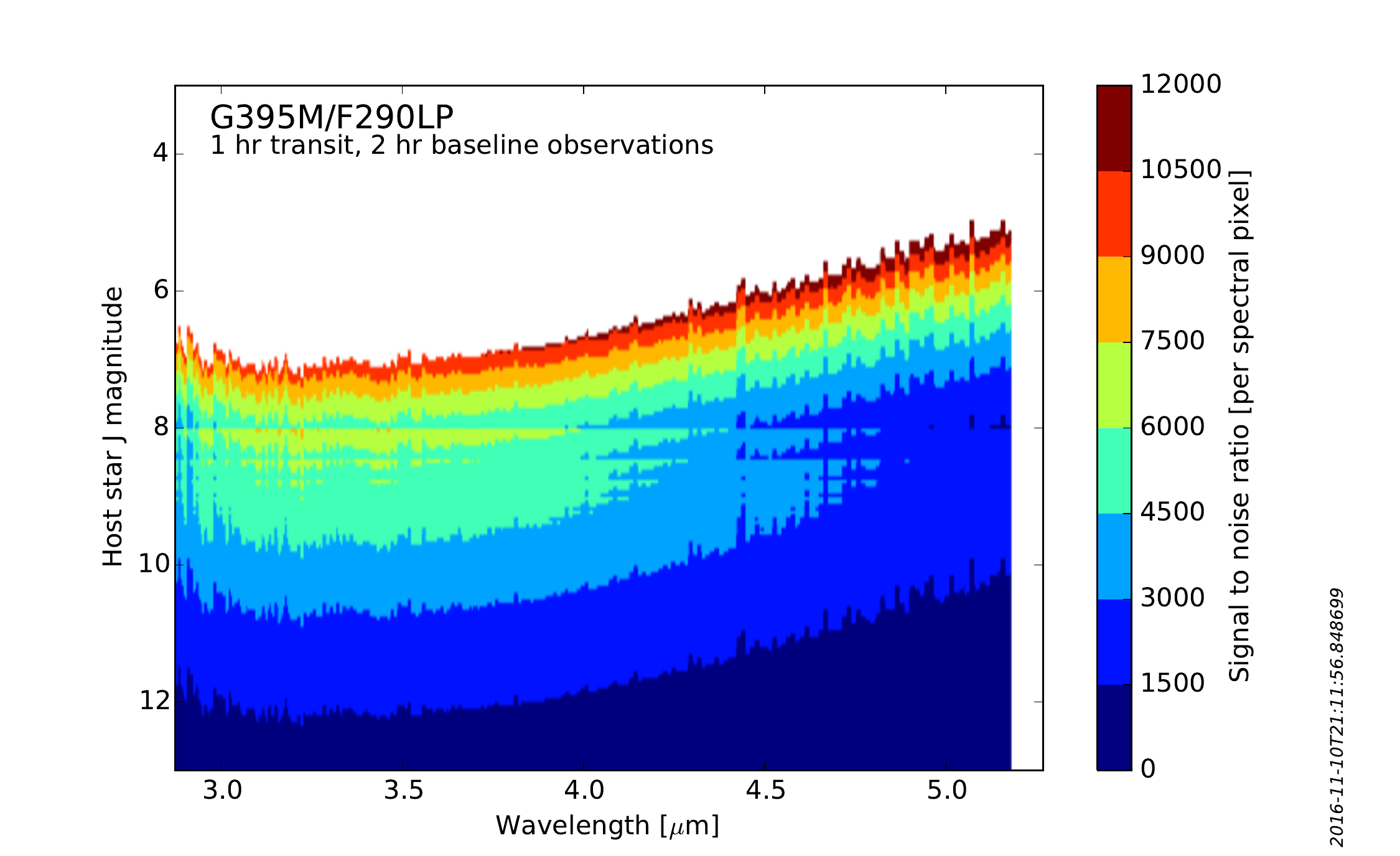}
\includegraphics[trim={1.3cm 0.2cm 1.3cm 0.5cm},clip,width=0.492\hsize]{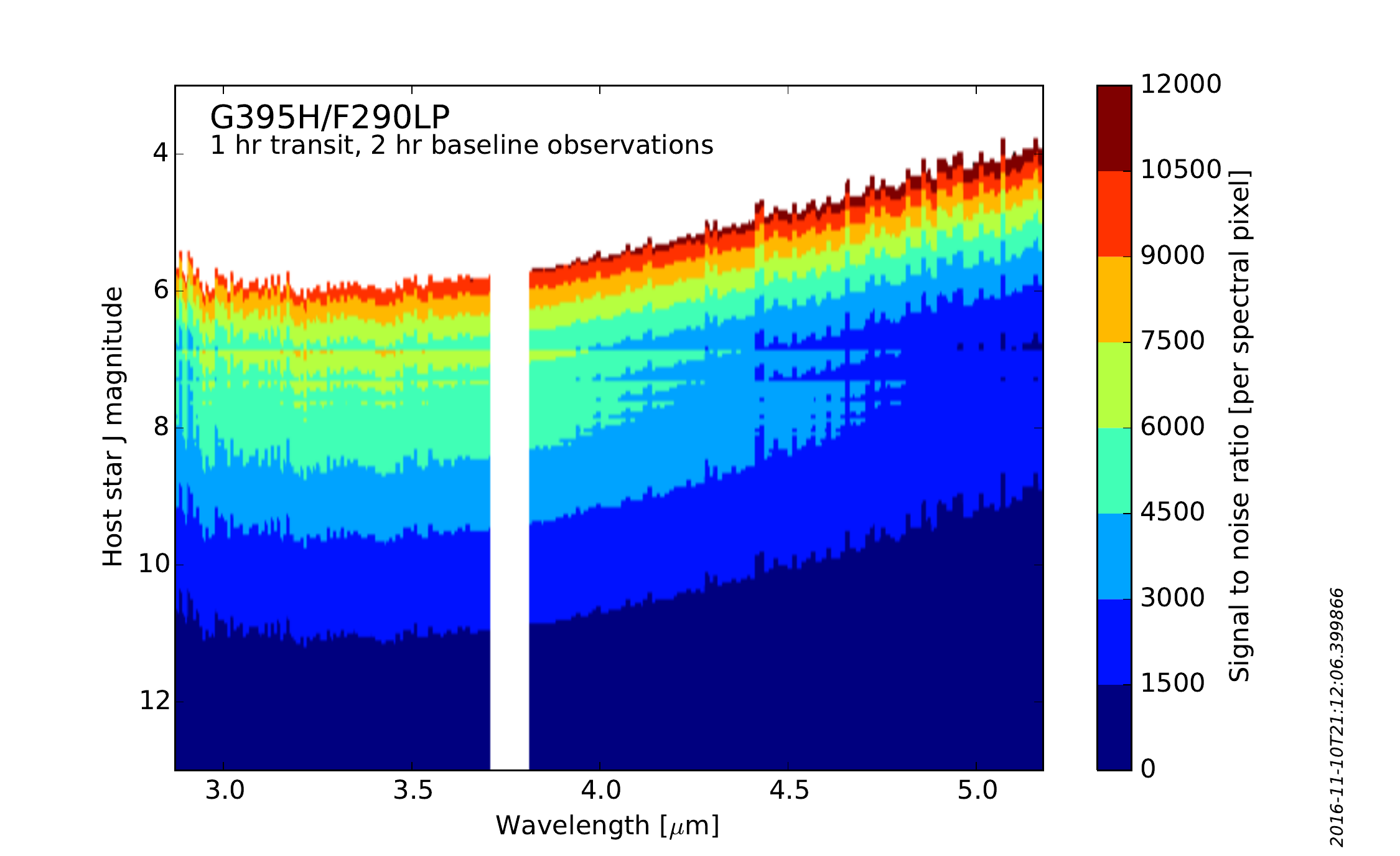}
\caption{\label{fig:snr_gratings}Signal-to-noise estimates for the high and medium spectral resolution gratings. Based on a host star with effective temperature $3400~\mathrm{K}$ and $\log(g)=4.5$ (cgs).}
\end{figure*}

Figure \ref{fig:snr_gratings} shows the S/N calculated in the same way for the gratings. The S/N for the gratings is much more uniform across the wavelength range than for the PRISM, as a result of the more uniform dispersion. An S/N of approximately 10,000-20,000 can routinely be achieved with the PRISM for transiting exoplanets, but is lower for the medium and high resolution gratings, at about 5,000-10,000. This is due to the shorter frame times of the smaller subarray that can be used for the PRISM, resulting in more integrations per transit and thus more photons collected. It is evident that the highest S/N is reached closest to the saturation limit and a difference of two in magnitude can double the S/N. Binning the grating-spectra in spectral direction to similar resolution as the PRISM will result in much higher S/N for these configurations. A closer comparison of the low, medium and high spectral resolution modes are presented in case studies in Section~\ref{sec:examples}.

We note that the details in the S/N figures will change for earlier-type stars due to the difference in spectral features. However, both the principal shape and the banding in S/N for different host star magnitudes will remain, as this is due to the instrument throughput and dispersion, and the change in efficiency of the readout and the resulting increase in photon collection time when an additional frame can be taken before saturation. We also point out that these estimates are based on ground test data acquired before launch. New in-orbit data will become available during commissioning and cycle 1 calibration, and as for the saturation limits observers are encouraged to check with the official JWST tools, that is, the JWST ETC and ExoCTK\footnote{\url{https://exoctk.stsci.edu}} \citep{Fowler2018}.

\section{Sources of systematic errors and their impact}
\label{sec:sys}

As shown in the previous Section, the performance of NIRSpec in terms
of the noise floor is excellent. There are, however, a number of
instrumental effects and calibration uncertainties that can increase
the actual noise level in NIRSpec observations. To assess the
stability of our system and the impact of systematic effects we
performed a dedicated test during the third cryo-vacuum campaign (henceforth CV3) of the JWST Integrated Science Instrument Module, that took place at
Goddard Space Flight Center during the winter 2015-2016
\citep{kvv+2016}, and acquired a 3-hour exposure of a bright
source in PRISM/CLEAR configuration.

As detailed in the Appendix, the analysis of this data set shows that
when observing a bright target for a time-scale typical of an
exoplanet transit, the dominant sources of systematic noise in the
NIRSpec data are the $1/f$-noise component from the detector
electronics and, in conjunction with significant pointing jitter and
drift, the signal variations in individual pixels due to an
undersampled PSF and the source movement.

The presence of correlated noise due to the readout electronics, that
manifests itself in low level cross-dispersion stripes and bands
visible in the count rate images, is well known in Teledyne HAWAII
detectors (see, for example, \cite{maf+2010}, \cite{raf+2012}, and also Paper I). In full
frame mode, reference pixels are used to track the drifts and minimize
the impact of this $1/f$-noise component. In the case of a subarray
exposures of a point source, there are several pixels
in each column that are not illuminated and can be used to track the
detector drifts and to effectively filter out this noise
component, as detailed in the Appendix. The analysis also identifies
the processing steps that can considerably reduce the impact of the
signal fluctuations arising from pointing drifts, and therefore
provides confidence in the stability of NIRSpec over the time scale
typical of a transit observation and the ability of the instrument to
reach the noise-floor of 200 parts per million (ppm) in less than five minutes of
integration.

\begin{figure}
\centering
\includegraphics[trim={3cm 0cm 1cm 0cm},clip,width=\hsize]{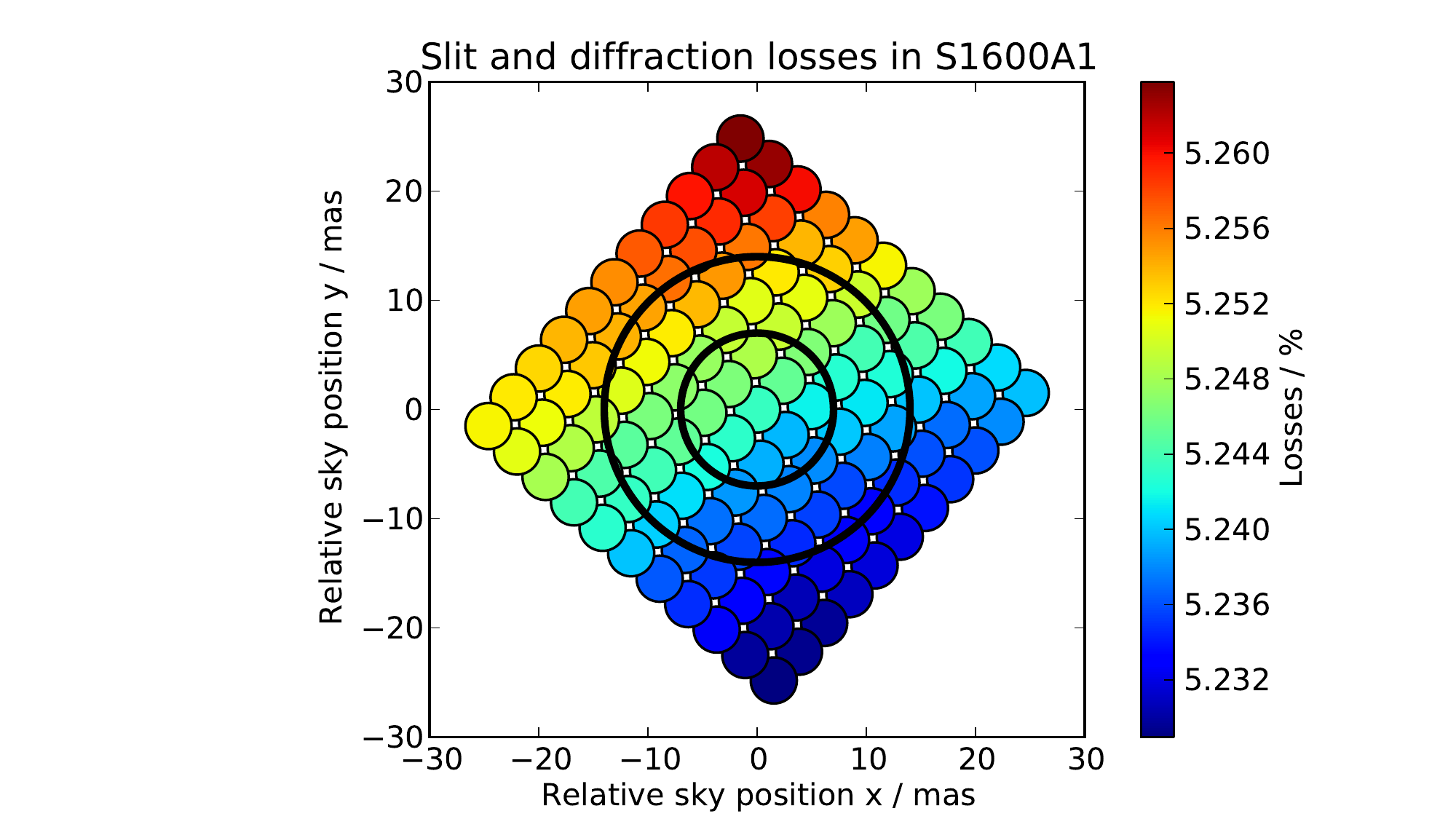}

\caption{\label{fig:sdlosses}Map of combined slit- and diffraction-losses in the central part of S1600A1 for the grating G235H at $2.45~\mu$m. The black circles have radius of 7 and 14 mas, corresponding to 1 and 2$\sigma$ of the allowed pointing drift over 10,000 s, as specified in JWST requirements.}

\end{figure}

Among other possible sources of systematic noise that could limit the
instrument performance in the observation of exoplanets, the
combination of pointing jitter and drift with intrapixel sensitivity
variations and variable slit- and diffraction-losses have been
identified as possibly the most significant.
The effects of this combination were evaluated using the NIRSpec Instrument
Performance Simulator (IPS) \citep{pll+2008, plp+2010}.
Fig.\,\ref{fig:sdlosses} provides a map of the combined flux losses due to
the telescope PSF being truncated and diffracted by the slit, for the
case of grating G235H. The typical level of intensity loss is $\sim$
5\%, with peak-to-valley variations of 0.03\% across the grid. The path-losses seen in orbit will depend on the actual PSF after the alignment and phasing of JWST during its commissioning.

The pointing stability of JWST is required to be better than 7 mas (1$\sigma$) over 10,000 s. This, however, includes random jitter over a time scale of seconds and longer term pointing drift that would appear as systematic signal drift, but would not directly contribute to the noise in a series of integrations that are short, typically a few seconds or less, depending on source brightness. Thus, to estimate the noise contribution in an exposure of exoplanets that consists of multiple integrations, we adopt the pointing requirements for NIRCam for timescales of a few seconds, which is 3.5 mas in line of sight and roll stability, corresponding to a jitter radius  of 5 mas (1$\sigma$). The relative noise in one exposure introduced by the telescope movement in a random direction is then the RMS of the throughput of S1600A1 on a circle of 5 mas radius relative to the value at its center, at the different wavelengths. The mean relative noise in an exposure of a point source over the wavelength range of the different dispersers are summarized in Table\,\ref{tab:sdlosses}. This noise contribution is expected to be below 25 ppm for all modes.

\begin{table}
\caption{Mean relative noise in an exposure due to slit- and diffraction losses.}
\label{tab:sdlosses}
\centering          
\begin{tabular}{lccc}
\hline\hline  
~ & ~ & Relative Noise $[10^{-5}]$ & ~\\
Resolution & Band I & Band II & Band III\\ 
\hline                    
R100 & ~ & 2.2 (over 0.6$-$5$\mu$m) & ~ \\
R1000 & 1.7 & 2.0 & 2.6\\
R2700 & 1.6 & 1.9 & 2.3\\
\hline                  
\end{tabular}
\tablefoot{\scriptsize Calculated for a point source in S1600A1 observed in the different bands, assuming a telescope random jitter of 5 mas.}
\end{table}

Of greater concern could be the interplay between the telescope
pointing instability and intrapixel sensitivity variations. Data on
characterizing the intrapixel sensitivity for these pixels are very
limited. \cite{hbp+2008} have characterized the intrapixel
sensitivity variations of an older-generation H2RG detector with a
resolution of approximately 4 $\mu$m (0.22 pixel) in the wavelength range
between 0.65 and 2.2 $\mu$m, for a sample of 64 pixels. The intrapixel
response is overall smooth with variations across the pixels at the
level of 2-3\% RMS, which is down to the measurement accuracy level,
however there appears to be some localized sensitivity dips and long
linear defects affecting roughly 10\% of the pixels
examined. \cite{hwp2014} completed a similar set of measurements on
the new design H2RG detector similar to the two detectors installed in
NIRSpec. They find intrapixel sensitivity variations with similar or
somewhat smaller amplitude than those from the earlier set of
measurements.

Using the intrapixel sensitivity maps from \cite{hbp+2008} convolved
with the NIRSpec PSF at 2 $\mu$m, we evaluated the interplay between
intra-pixel sensitivity and telescope pointing variations (RMS $< 7$
mas), by drawing 1000 random samples around each subpixel within an
area with 1$\sigma$ radius of 6.5 mas and then computing the $1\sigma$ over the
1000 samples.  We found the typical $1\sigma$ of this noise component
to be $\sim$400 ppm of the signal amplitude, if uncorrected for (and if
the small sample of pixels available is reasonably representative). In
this respect, therefore, it is encouraging to see from our
analysis of the CV3 time-series that a simple bi-linear fit of the
source spatial jitters and long term (large) drift that corrects for
the PSF undersampling (and potentially also corrects for
intrapixel sensitivity variations) allows us to reach noise-to-signal (N/S) below 200
ppm. Note that the 5-color light curve division applied to our data set (see Appendix~\ref{app:a} for details) would not be able to correct for the effect of intrapixel sensitivity variations, as this noise component acts strictly at the
level of individual pixels.

Another instrumental systematic behavior that could affect NIRSpec
observations of bright targets, and that cannot be probed with the CV3
data (due to the light source instability), is detector persistence.
Persistence refers to the detector's ``memory'' of previous exposures. It is one manifestation of charge trapping. Charge traps are localized defects in the detector that present unintentional potential wells that can temporarily capture mobile charges. If a moving charge falls into a trap before being integrated, it will become stuck and will not appear as part of the signal in that exposure. When the trap decays and releases the charge at a later time, it becomes mobile again. Most of the time, the newly released charge will integrate along with the photocurrent from the current exposure to appear as persistence.

The effect of persistence when observing bright host stars with Spitzer or HST has been reported by several authors, for example, \citet{ckb+2008}, \citet{ack+2010}, \citet{bcde+2012}, and \citet{wse+2016} and is referred to as the ``ramp'' or ``hook''
effect, because of the shape of the light curve over a set of contiguous exposures: sharply rising and quickly saturating ramp-like features, as can be clearly seen in, for instance, the observation with HST/WFC3 in stare mode of GJ1214 by \citet{bcde+2012}. The ``charge trapping'' model leads to exponential ramps when observing bright sources as the excess dark current increases sharply at first, but slows its increase as the population of charge traps begins to approach steady state.

NIRSpec detectors are also affected by persistence, and as exoplanet observations will be conducted without dithers within slit S1600A1, we can expect a detector behavior qualitatively similar to that of HST/WFC3 when used in stare mode, that is, an initially rising light curve during the first integrations of a bright target, followed by an equilibrium state in which flux levels are highly repeatable. Quantitatively speaking, however, the effect is expected to be significantly smaller for the much newer NIRSpec detectors. 

\citet{Rauscher+2014} have characterized the persistence behavior of the two NIRSpec detectors. They measured persistence in dark exposures acquired soon after illuminating the detector with flat-field light at a fluence of 10 $\times$ the full well level and found that the persistence of NIRSpec flight SCAs was well approximated by a power law, $f(t)=f_0 (t/1~{\rm s})^{-1}$ with $f_0$ equal to 29 and 42 ${\rm e^-  s^{-1} pix^{-1}}$, respectively for NRS1 and NRS2. In comparison, \cite{mcd+2010} used data from persistence tests of WFC3 IR flight array acquired at the Detector Characterization Lab at Goddard Spaceflight Center by \cite{hilbert2008} using 100 times full well overexposed flat field illumination to show that the persistence decay in WFC3 detector follows a power law of the form 0.156 $(t/1~{\rm hour})^{-1}$. This corresponds to a latent signal of $\sim 5.6$ ${\rm e^- s^{-1} pix^{-1}}$ after 100\,s since the end of the over-exposure, compared to a corresponding signal from NIRSpec's detectors of less than 0.5 ${\rm e^- s^{-1} pix^{-1}}$. Note that in the case of the NIRSpec detector arrays, using ten times or 100 times fluence level in the over-exposure does not impact the latency level, as persistent images do not get stronger over $\sim 2-3 \times$ saturation. Also, persistence does not depend on the wavelength of the illumination, but does vary with illumination time, in the sense that the longer the over-illumination, the stronger the persistence image \citep{Rauscher+2014}.

We also note that in the case of NIRSpec, up to 65,535 integrations can be taken during one (hours-)long exposure without breaks, because, unlike for HST's orbit, JWST's position at L2 enables its instruments to perform continuous observations of a target for many hours or even days (see Section~\ref{sec:planning}). This will allow observers to maximize the time the detectors will be in their equilibrium state, for example, by having a relatively long baseline before the event, which in turn should minimize the effect of charge trapping for transit, occultation, and phase curve observations.


\section{Case studies} \label{sec:examples}
We present simulated observations based on synthetic spectra of the well studied exoplanets HAT-P-26\,b \citep{Hartman2011} and GJ\,436\,b \citep{Butler2004,Gillon2007}, and the radiometric model utilized in Section~\ref{sec:SNR}. All examples are based on one transit and one occultation with twice the transit or occultation duration spent on baseline observations. Simulations were done using a set of python routines and using the instrument throughput and detector noise performance, but do not include the potential impact of the systematic errors discussed in Section~\ref{sec:sys}.

The calculation of the synthetic spectra is based on radiative transfer models developed and tested in the context of both solar system and exoplanet atmosphere investigations \citep[see e.g.,][]{Munoz2012,Munoz2013,MunozIsak2015}. The spectra for reflected and thermal emission are calculated at a star-planet-observer phase angle of zero, and integrate the signal over the whole planet disk. 

We estimate the planet equilibrium temperature as $T_{eq}=T_{\mathrm{star}}\left( R_{\mathrm{star}} / 2a \right)^{1/2}$, where $T_{\mathrm{star}}$ is the stellar effective temperature, $R_{\mathrm{star}}$ is the stellar radius and $a$ is the orbital semi-major axis. For simplicity, we parameterize the temperature such that it is constant and equals to $T_{eq}$ from 1,000~bar to 0.1~bar. At higher altitudes, the implemented temperature drops exponentially with a scale pressure of 0.1~bar toward the limiting value of $2/3~T_{eq}$ at the atmospheric top. This limiting value is roughly consistent with the temperature profiles without thermal inversions that appeared in the literature for this planet \citep{Hu2015}. Most of the thermal radiation shown in the corresponding emission spectra traces the equilibrium temperature of the planet, and therefore the choice for the limiting value at higher altitudes is not critical. For GJ\,436\,b, we adopted the following altitude-independent volume mixing ratios: \chem{0.17 ~(He}); \chem{9.1x10^{-4} ~(H_2O)}; \chem{4.9x10^{-4}~ (CH_4)}; \chem{1.2x10^{-4}~( NH_3 )}, with the rest as \chem{H_2}. Correspondingly, for HAT-P-26\,b: \chem{0.17 (He)}; \chem{2.4x10^{-7}~ (K)}; \chem{2.96x10^{-6} ~(Na)}; \chem{4.2x10^{-4} ~(H_2O)}; \chem{6x10^{-5}~ (N_2)}; \chem{4.9x10^{-4} ~(CO)}. These volume mixing ratios were chosen on the basis of solar elemental abundances and assumed that only the listed gases are present in the atmosphere. For a cool planet such as GJ 436b, it is expected that \chem{H_2O}, \chem{CH_4} and \chem{NH_3} take most of the \chem{O}, \chem{C} and \chem{N} atoms. In turn, for a hotter planet such as HAT-P-26b, it is expected that \chem{H_2O}, \chem{CO} and \chem{N_2} play an equivalent role. Further, we assumed that the alkalis have precipitated deep in the atmosphere of GJ 436b and their signature is not discernible on that planet. The spectroscopic information for the gases is taken from HITEMP \citep{Rothman2010} for the \chem{H_2O} molecule, and from HITRAN2012 \citep{Rothman2013} for the other molecules. Absorption by the sodium and potassium atoms is parameterized as in \cite{Iro+2005}. Temperature-dependent absorption due to \chem{H_2-H_2} and \chem{H_2-He} is included, with the corresponding cross sections from HITRAN2012.

The atmospheric models consider the effect of haze. The baseline implementation, thin haze, considers number densities that are equal to those of the gas. The variation thick haze considers aerosol densities one hundred times those in the baseline model. For the haze particles, a wavelength-dependent cross section of $10^{-28}~(1~\mathrm{\mu m} / \lambda )^2 ~\mathrm{[cm^2]}$ is assumed. This cross section is comparable to that for Rayleigh scattering at wavelengths near $1~\mathrm{\mu m}$. We assumed that the haze particles scatter conservatively with a single scattering albedo of one. The planet's geometric albedo is not prescribed but results from the solution of the multiple scattering problem.

\subsection{HAT-P-26\,b}
The S/N calculations for one transit of the low density Neptune-mass planet HAT-P-26\,b is based on the stellar and orbital parameters presented in \citet{HAT-P-26b}. A PHOENIX model spectrum with $T_{\mathrm{eff}} = 5100~\mathrm{K}$, $\log(g)=4.5$ and magnitude 10.80 in the J band has been adopted to represent the K1 dwarf host star. A transit duration of $8800~\mathrm{sec}$ is used in the simulations, resulting in a total exposure time of 7hr20min for one transit. This includes overheads due to reseting the detector between integrations, but not pad time where the system has time to stabilize or other parts of the visit where the telescope and instrument are setup for the observation.

\begin{figure*}
\includegraphics[trim={2.1cm 0cm 2.1cm 0cm},clip,width=\hsize]{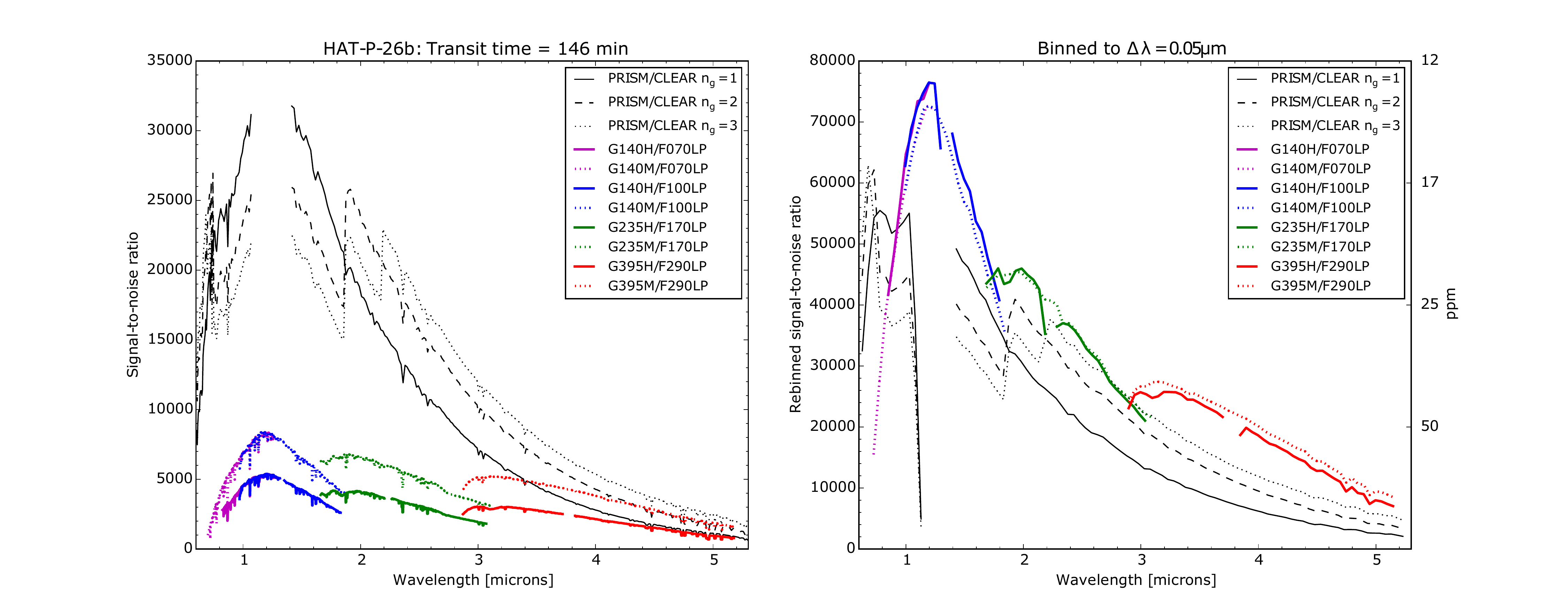}
\caption{Signal-to-noise ratio for one transit observation (2hr26min) of HAT-P-26\,b, spending twice that amount of time on baseline out-of-transit observations. Left panel shows the S/N for the native instrument resolution, and the right one is binned in spectral direction to a dispersion of $\Delta \lambda = 0.05 ~\mathrm{\mu m}$. Calculations are based on a host star with $T_{\mathrm{eff}} = 5100~\mathrm{K}$, $\log(g)=4.5$ and J magnitude 10.80.}
\label{fig:snrHAT}
\end{figure*}

The S/N curves presented in Fig. \ref{fig:snrHAT} allow for direct comparison between all of the available configurations, both in their native instrument resolution (left panel) and spectrally rebinned to a uniform dispersion of $\Delta \lambda = 0.05~\mathrm{\mu m}$ (right panel). The solid, dashed and dotted black lines represent the low resolution configuration using one, two, or three groups per integration ($n_g$) respectively. The missing wavelength-coverage at $1-1.4~\mathrm{\mu m}$ is due to partial saturation of the spectrum already in the first group. The S/N will in some wavelength ranges go up when increasing the number of groups per integration, whereas in other regions of the spectrum the detector will saturate in group two or three, leading to a lower duty cycle as these groups will be discarded in the data processing. For the PRISM/CLEAR configuration it can be advantageous to use $n_g=3$ in order to get better S/N red-ward of $2~\mathrm{\mu m}$.

From the S/N curves which have been binned in the spectral direction, it is clear that the high and medium resolution options result in similar S/N when binned to the same dispersion. It will therefore in most cases, when the spectral resolution is not needed, be of interest to use either the low or medium resolution configurations for objects which are not too bright for these modes in order to maximize wavelength coverage.

\begin{figure}
\includegraphics[trim={0.5cm 0.5cm 1.5cm 0cm},clip,width=\hsize]{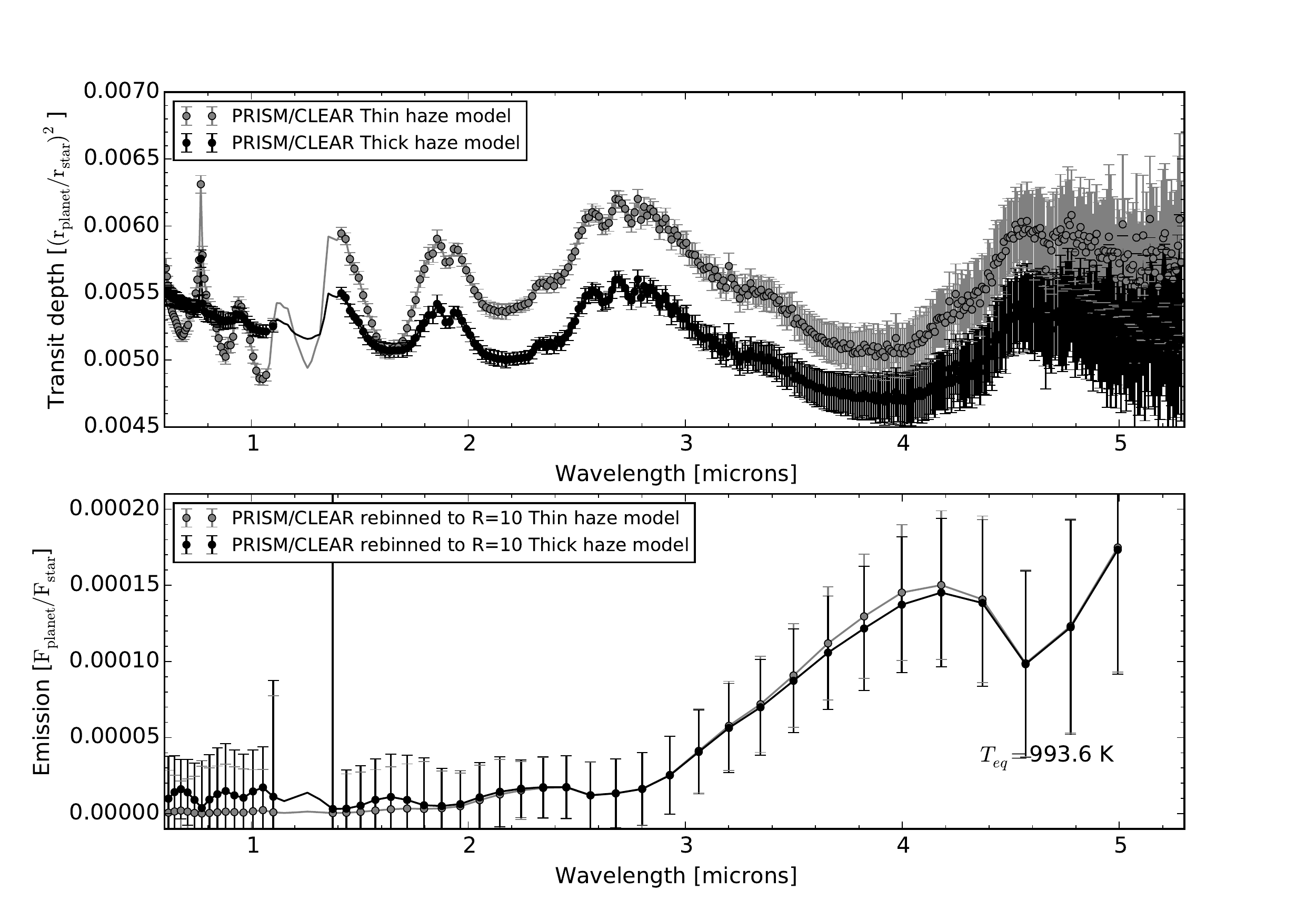}
\includegraphics[trim={0.5cm 0.5cm 1.5cm 1.5cm},clip,width=\hsize]{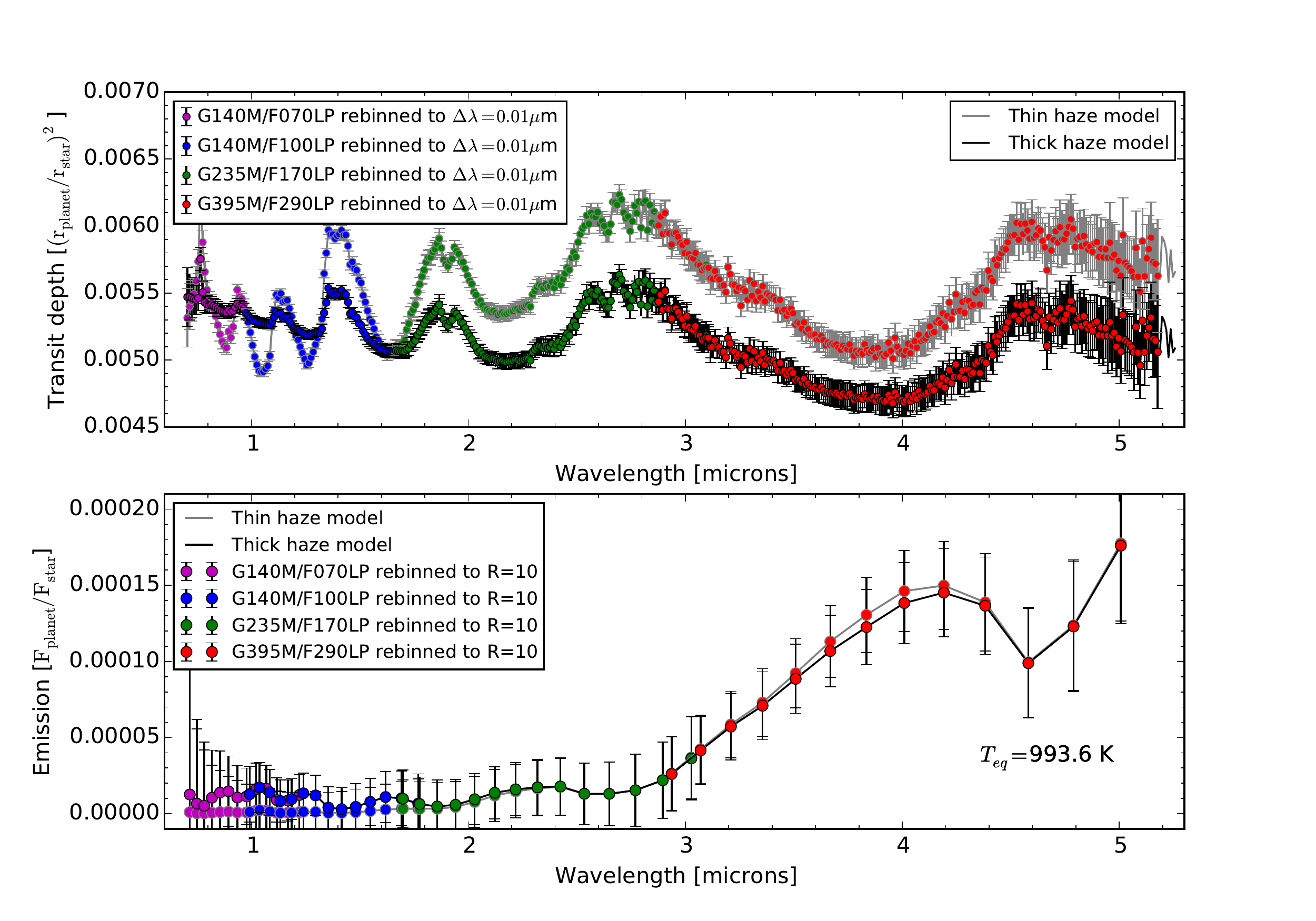}
\caption{Modeled transit depths and emission spectra for HAT-P-26\,b, with the equilibrium temperature assumed in the atmospheric model indicated in the panels showing emission. The dots and curves represent thin and thick atmospheric models (disperser used is color coded), and the error bars give the expected 1$\sigma$ uncertainties for NIRSpec transit and occultation measurements after rebinning to the indicated resolution or $\Delta\lambda$ (photon and read noise only).}
\label{fig:hatp26b}
\end{figure}

Moving on with only the medium resolution configuration plus the low resolution PRISM with $n_g=3$, the planetary signal has been simulated using the atmospheric planet models as a contrast to the host star for both the thin and thick haze cases. Figure \ref{fig:hatp26b} show transit depth, $(F_{out}-F_{in}) / F_{out} =$ $( r_{\mathrm{planet}} / r_{\mathrm{star}} )^2$, and emission ($(F_{\mathrm{thermal}}+F_{\mathrm{reflect}}) / F_{\mathrm{star}}$) consisting of both thermal emission, $F_{\mathrm{thermal}}$, from the planet itself and reflected star light, $F_{\mathrm{reflect}}$. The two gray curves represent the thin and thick atmospheric haze models and error bars correspond to the expected 1-$\sigma$ uncertainties of NIRSpec measurements of one transit or occultation, respectively. 

Distinguishing the thin and thick haze cases can easily be done with the low resolution configuration for the transit, even without further binning of the spectrum. In emission the data-points are rebinned to $R=10$, and the two models are hard to separate, as both haze configurations become transparent to long wavelength radiation. Around $1~\mathrm{\mu m}$ the two models differ slightly due to reflected star light, but on a scale that cannot be measured in one occultation. 

In order to cover the same wavelength range with the medium resolution gratings, four transits or occultations must be observed. The result is a trade-off between higher spectral resolution and/or better S/N than the low resolution PRISM. The two lower plots in Fig. \ref{fig:hatp26b} show the transit and occultation with the four medium resolution grating and filter combinations, for the transit depth rebinned to a uniform dispersion of $0.01~\mathrm{\mu m}$ and the occultation rebinned to spectral resolution R=10, enabling direct comparison with the PRISM-measurement.

In the case of HAT-P-26\,b there are clear advantages in using the PRISM/CLEAR configuration for the observations, whether it is transit or occultation, as the low resolution mode gives the widest wavelength coverage and S/N comparable to the medium resolutions gratings rebinned. However, if a science case requires resolution higher than 100, extremely high S/N (equivalent to better than 40 ppm) or guaranteed wavelength coverage at $1-1.4~\mathrm{\mu m}$, one of the medium modes might be the preferred option.

\subsection{GJ\,436\,b} 
From Fig. \ref{fig:sat2} it is clear that the host star of exoplanet GJ 436b with is J-band magnitude of 6.9\,mag is too bright to be observed with the PRISM. Further analysis of the observability and S/N is done by using a PHOENIX template with $T_{\mathrm{eff}}=3700 ~\mathrm{K}$, $\log(g)=5.0~\mathrm{(cgs)}$ and J~magnitude 6.90 \citep{Torres2007,2MASS}. More than half of the spectrum is saturated when using medium resolution gratings, leaving the high resolution modes as the most suitable configurations for GJ\,436\,b.

\begin{figure}
\includegraphics[trim={0.5cm 0.5cm 1.5cm 0cm},clip,width=\hsize]{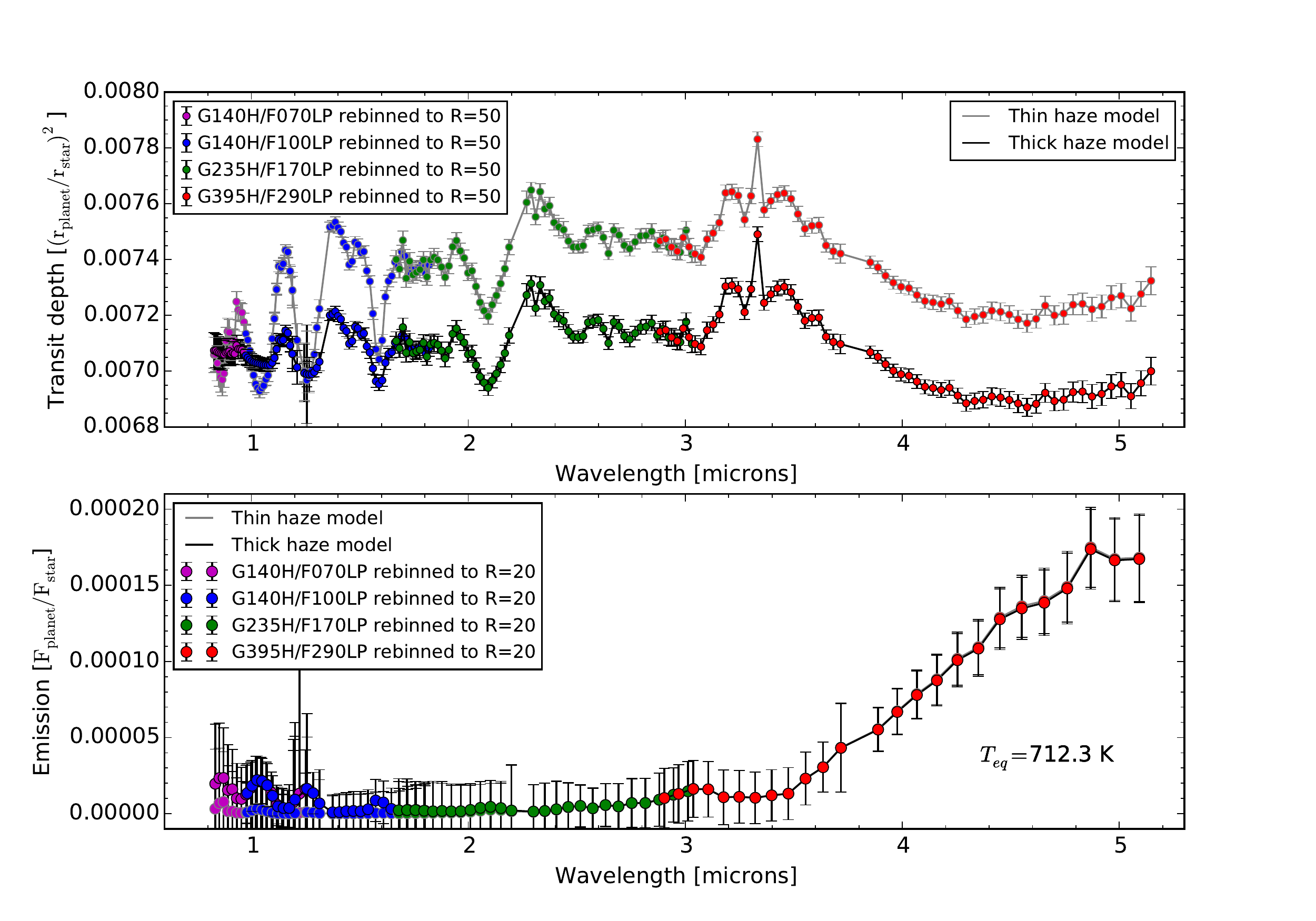}
\caption{Modeled transit depths and emission spectra for GJ\,436\,b. The gray curves represent thin and thick atmospheric models and the error bars denote the expected 1$\sigma$ uncertainties for NIRSpec transit and occultation measurements with the high-resolution gratings, after rebinning to the indicated resolution (photon and read noise only).}
\label{fig:gj436b}
\end{figure}

Figure \ref{fig:gj436b} shows our modelled transit depths and emission spectra with error bars as expected when using the high resolution gratings, based on an eclipse of duration 2850 seconds \citep{Morello2015}. Getting the full wavelength coverage from 0.8 to $5.2~\mathrm{\mu m}$ takes four eclipses, as only one configuration can be observed at a time. The number of groups per integration have been tweaked for the two configurations covering the wavelength longward of $2~\mathrm{\mu m}$ in order to get better S/N in the red parts of the spectra. This means sacrificing a bit of data quality in the blue of each individual spectrum, as was done for the low resolution mode in the case of HAT-P-26\,b. This procedure pays off the best when dealing with targets that are close to the saturation limit of the given instrument setup.

With the transit data which has been rebinned to spectral resolution R=50, the two atmospheric models can easily be distinguished. In particular the G140H/F100LP could be a good choice of observation mode for the transit, as it probes the wavelengths most sensitive to the atmospheric composition. Thermal emission from the planet can be detected with G395H/F290LP rebinned to R=20 for the occultation.

\section{Planning BOTS observations} \label{sec:planning}

As for all JWST programs, the Astronomers Planning Tool (APT)\footnote{The most recent version of APT can be found on the STScI website at \url{http://www.stsci.edu/hst/proposing/apt}} will be used to design NIRSpec observations. NIRSpec bright object time-series observations has its own template in APT. After selecting the NIRSpec instrument, the BOTS template, and the target, one will have to provide additional information in order to plan the observations. Those inputs are briefly described in the following subsections. Up to date information on the BOTS observing template can also be found on the STScI APT JDOX pages\footnote{\url{https://jwst-docs.stsci.edu/jwst-astronomers-proposal-tool-overview}}.

\subsection{Target acquisition}

Before the time-series spectra can be taken, the target has to be placed into the S1600A1 aperture. The 1$\sigma$ radial blind pointing accuracy of JWST is expected to be 0.15\,arcsec, with a stability on the order of 7\,mas\footnote{\url{https://jwst-docs.stsci.edu/jwst-observatory-characteristics/jwst-pointing-performance}}.  Centering a transit target star to better than 0.1\,arcsec precision is likely desired to limit any photometric effects from the pointing stability, which can approach the order of 10$^{-4}$ if the star is positioned beyond $\sim$0.1\,arcsec from the aperture center. Therefore, the observer can also opt to perform a target acquisition (TA) using the science target itself or a nearby (the allowed slew distance within a visit varies between 30" and 80" depending on Galactic latitude) reference source, which should be compact. NIRSpec will be used to take an image of the TA source and the on-board scripts will calculate the centroid and call for a small angle maneuver (SAM) to center it inside the S1600A1 aperture. In case the TA target is not the science target, this will be followed by a SAM to move the latter into the center of S1600A1.

The STScI exposure time calculator should be used to determine the best settings for the target acquisition. The proposer will have to select the target acquisition filter (available choices are F110W, F140X, and CLEAR), readout mode (NRSRAPID or NRSRAPIDD6), and subarray (\verb+FULL+, \verb+SUB2048+, ]\verb+SUB32+) in order to achieve sufficient signal-to-noise (10 to 20) on the TA target without saturating it (see Paper I for more details). 

\subsection{Instrument configuration and exposure setup}

The BOTS observing template in APT also contains a ``Science Parameters'' section where one can specify the subarray and grating/filter combination. Tables \ref{tab:subarrays} and \ref{tab:modes} list the available choices which will depend on the science case, for example, the desired resolution and wavelength coverage.

The proposer also has to set the readout pattern (NRSRAPID or NRS), number of groups per integration, number of integrations and the number of exposures. One should always make use of the ETC to determine how many groups per integration can be taken before detector saturation occurs. As discussed in Section~\ref{sec:detector}, the NRSRAPID readout pattern is preferred for bright targets, because it allows for more groups before saturation than the NRS mode. The higher the number of groups, the better the duty cycle and thus overall efficiency (see Section~\ref{sec:SNR}), due to the time needed to reset the detector. Therefore, one should make the number of groups as large as possible without saturating the detector, unless shorter integrations are desired, for instance, when a better time resolution is required.

Finally, the number of integrations and exposures are set. They determine --- together with the choice of subarray, readout pattern and number of groups discussed before --- the total length of the observation. For primary and secondary eclipses, this total time will typically be the transit duration plus some baseline before and after the event. Whenever possible, integrations (up to 65,535) should be used to get to the desired duration. This is due to the fact that the NIRSpec detectors return to full-frame idle mode after the completion of each exposure and this will introduce thermal transients in the detector and readout electronics that can affect the radiometric stability of the NIRSpec detector chain. If that is not possible (e.g., for long phase curve observations of very bright targets) and more than one exposure is necessary, one should make sure that the exposure break does not happen at a critical time during the event.

Dithers are not supported in the BOTS observing template, because this would be detrimental for the spectrophotometric stability due to the fact that i) the observation would need to be broken up into several exposures (per dither position) which is not advisable (see above) and ii) the source position affects slit and diffraction losses and the relative flat fielding accuracy becomes important and potentially limiting when different pixels are illuminated by the (undersampled) PSF.

\subsection{Timing constraints} \label{sec:timing_constraints}

While in general all JWST observations will be event driven, observers can provide timing constraints for time critical observations in APT. The absolute timing of JWST NIRSpec observations of exoplanet eclipse or phase curve targets will be controlled by the following special timing requirements: the Julian Date of the zero phase of the event (e.g., the transit or eclipse), the period of the planet's orbit, the earliest allowed starting time (phase) of the observation, and finally the latest allowed starting time (phase) of the observation.
In APT, these are specified as a ``Timing->Phase'' special requirement and APT will use these constraints when calculating the scheduling windows for visit planning. Timing constraints make scheduling observations more challenging and can have a negative impact on overall observatory efficiency due to imposed waiting times in the event driven flow. Therefore, a ``direct scheduling overhead'' of 1 hour will be added onto programs that have a tightly constrained timing requirement (start window less than 1 hour wide). The shortest absolute timing constraint for JWST is five minutes, and the window defined by the earliest and latest phase start time must be equal or larger than that.

Due to the thermal settling time of the detectors that will happen at the beginning of the exposure, it is advisable to allow for extra ``padding'' of about 15 minutes at the beginning of the observation by adjusting the phase range accordingly.

\subsection{High gain antenna moves}

During normal JWST operations, the High Gain Antenna (HGA) must re-point occasionally to keep continual contact with ground communication stations. These HGA moves are estimated to occur every $\sim$10,000\,s, and they take approximately one minute to execute. During the HGA moves, the guide star lock is dropped out of its fine lock guiding mode, which can cause temporary deviation movement of the guide star (and hence science target) positioning. Therefore, the maximum exposure duration is limited to the 10,000\,s HGA re-point cadence for most of JWST's observing cases, and the antenna moves will only happen between the exposures. For exoplanet transit and occultation observations, however, the expected exposure times necessary to monitor a full eclipse light curve can often far exceed the 10,000\,s limit, and exposure breaking around HGA moves would be difficult to implement in an optimal way that would not affect the light curve stability. Therefore, the exposure duration in the NIRSpec BOTS template is only limited by the choice of subarray and number of groups (and ultimately the 48 hour visit duration limit) and exposures will continue through the HGA re-point moves. Because these observations are a time series, there will be some integrations within the light curve that might be affected by pointing instability induced by the HGA moves, as the guider drops the lock. After the HGA moves, the guide star lock and repositioning is expected to be repeatable to better than 5 mas, and therefore should not significantly alter the transit light curve signal outside of the move time.

\subsection{BOTS observation flow}

\begin{figure}
\centering
\includegraphics[width=\hsize]{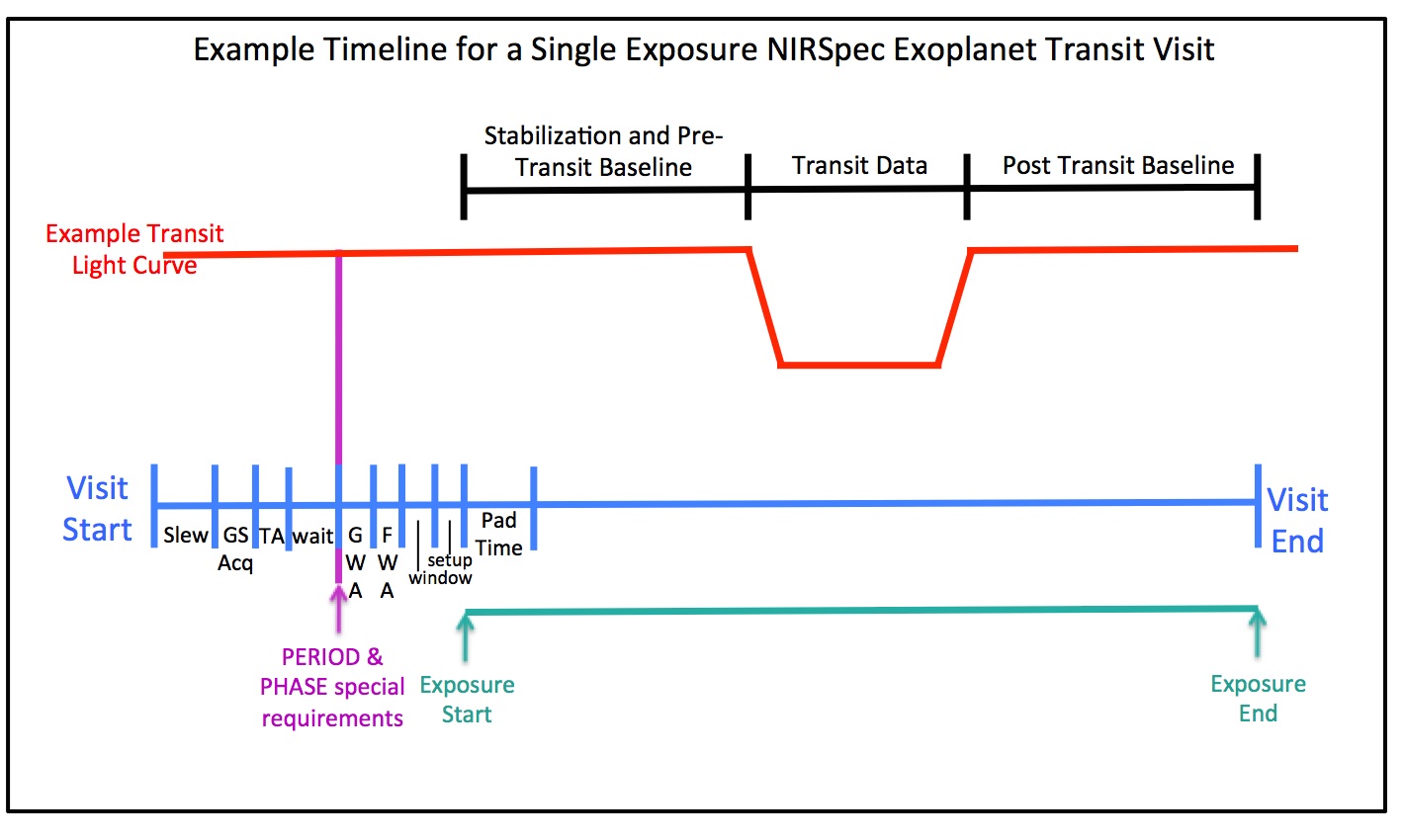}
\caption{\label{fig:transit_visit}Example timeline for a single exposure NIRSpec exoplanet transit observation. Times represented in the visit timeline are not to scale.}
\end{figure}

Figure~\ref{fig:transit_visit} presents the sequence of activities in a single exposure NIRSpec exoplanet transit observation. The visit starts with the slew to the transit star, followed by the guide star (GS) acquisition and the (optional) instrument target acquisition to place the star within the NIRSpec S1600A1 aperture. Depending on the timing of the visit, there might be a brief wait period prior to the configuration of NIRSpec for the first science activity. The start of the first science activity is seen as the magenta line and arrow in Fig.~\ref{fig:transit_visit}, and is defined by the phase range, period, and zero phase special requirements (see Section~\ref{sec:timing_constraints} above). The duration of the initial science activities to configure the instrument, for example, setting the filter wheel assembly (FWA) and the grating wheel assembly (GWA) for the first exposure, is expected to be less than about five minutes. The first (or only) exposure starts and some thermal settling will happen while data is being acquired in the exposure. In case of only one exposure it will continue uninterrupted for the duration of the visit. If there is more than one exposure they will be taken back-to-back, but there will be a short ($\sim$20\,s) gap between them due to the transition to full-frame-idle discussed above, which will also introduce small thermal transients. After the exposure(s) are complete, brief close-out activities (moving the FWA to its closed position) will occur at the end of the visit.

\section{Data pipeline for time-series observations} \label{sec:pipe}

\begin{figure*}
\centering
\includegraphics[width=0.9\hsize]{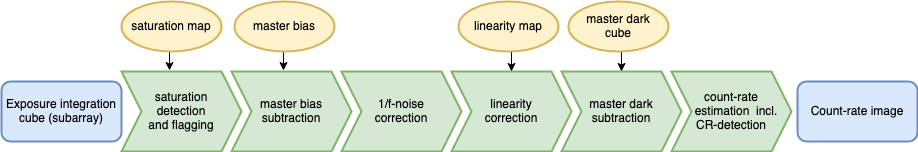}
\includegraphics[width=0.9\hsize]{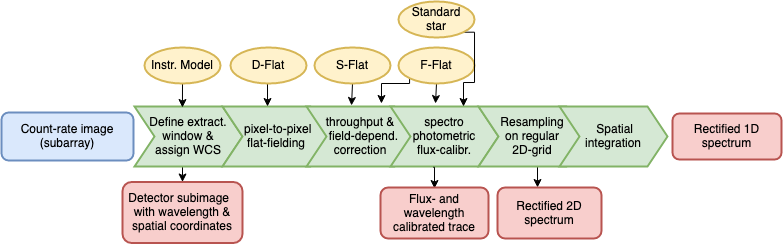}
\caption{\label{fig:pipeline} Default ramps-to-slopes (top) and slopes to calibrated data (bottom) processing steps for time-series observations.}
\end{figure*}

The data reduction steps for time-series observations (TSOs) are predominantly the same as for the ``standard pipeline'' that is used for other NIRSpec data. The overall approach to reduce and calibrate NIRSpec data has been described by \citet{rawle2016} and \citet{oliveira2018} and is also summarized in Papers II and III. The data reduction steps for TSO data are outlined in Fig.~\ref{fig:pipeline}. Due to the use of subarrays and the resulting lack of detector reference pixels for some of these, a dedicated $1/f$ detector noise correction will be performed using the data of pixels that do not see light, because they are outside of the spectral trace. The up-the-ramp fitting (count rate estimation) of the integrations also features a jump detection step for cosmic ray outlier rejection (see also \cite{Giardino_2019} for more details). There will be a count rate map for each individual integration that is then calibrated (wavelength solution, flat fielding, etc.) and the 1D spectrum extracted. A white light curve (summing up all wavelengths) will be created from the time-series data for quick look purposes.

Successful proposers will also be able to download the raw (up-the-ramp) data in order to be able to run their own data pipeline and analysis software when deemed necessary, and more details on and the most recent implementation of the TSO data reduction pipeline are available online via the ``James Webb Space Telescope User Documentation'' pages hosted by STScI.\footnote{\url{https://jwst-docs.stsci.edu/jwst-science-calibration-pipeline-overview/stages-of-jwst-data-processing}}

\section{Conclusions} \label{sec:con}

In this paper we highlight the capabilities of NIRSpec in observing bright host stars of transiting exoplanets via bright object time-series (BOTS) observations. Given the excellent throughput of NIRSpec and its relatively high spectral resolution, in combination with the stability and large collecting area of JWST, we can expect breakthrough discoveries in the field of exoplanet characterization. We show example case studies of planets HAT-P-26\,b and GJ\,436\,b to emphasis what can be done and achieved in principle, but will require careful calibration and mitigation of potential systematic errors also discussed in this paper.

We also mention that NIRSpec can be used to obtain spectra of compact directly imaged exoplanet systems such as TWA\,27 and HR8799 with its IFU. With its good image quality and spectral slices well-separated on the NIRSpec detectors, significant contrast ratios should be achieved. The same spectroscopic modes and wavelength coverage as for BOTS observations are also available when observing with the IFU. Details on integral field spectroscopy (IFS) with NIRSpec are given in Paper III.

\begin{acknowledgements}
NIRSpec owes its existence to the dedication and years of hard work of a great number of engineers, scientists and managers scattered across European and US industry, academia, and the ESA and NASA science programs. The contributions of these colleagues and institutions are too numerous to list here, but are greatly appreciated all the same.
L.D.N thanks the Swiss National Science Foundation for support under Early Postdoc.Mobility grant P2GEP2\_200044. AJB acknowledges funding from the ``FirstGalaxies'' Advanced Grant from the European Research Council (ERC) under the European Union's Horizon 2020 research and innovation programme (Grant agreement No. 789056).
We also thank the anonymous referee for the constructive and timely comments that have improved the presentation of this paper.
\end{acknowledgements}

%
\bibliographystyle{aa} 
\bibliography{thebibliography} 
%


\begin{appendix}
  
\section{Mimicking the observation of an exoplanet transit during ground testing}
\label{app:a}

To evaluate the response of NIRSpec to an exposure of a bright target
over a time scale typical of an exoplanet transit, we acquired a
$\sim$3 hour-long exposure of a bright source. That data was obtained during the third cryovacum campaign (CV3) of the JWST Integrated Science Instrument Module (ISIM), that took place at Goddard Space Flight Center (GSFC) in winter 2015-2016 \citep[][]{kvv+2016}. The exposure was taken with NIRSpec at the operating temperature of $\sim$40 K, using the PRISM, and consists of 12,000 integrations of three groups (of one
frame) each, in detector window mode, with subarray \verb+SUB512+ and an effective integration time of $t_{\rm int} = 0.45$ s per integration. The count rate image of the dispersed source in one integration is shown in Fig.\,\ref{fig:data_prism}.

\begin{figure}[b]
\includegraphics[trim={0cm 0cm 0cm 1cm},clip,width=\hsize]{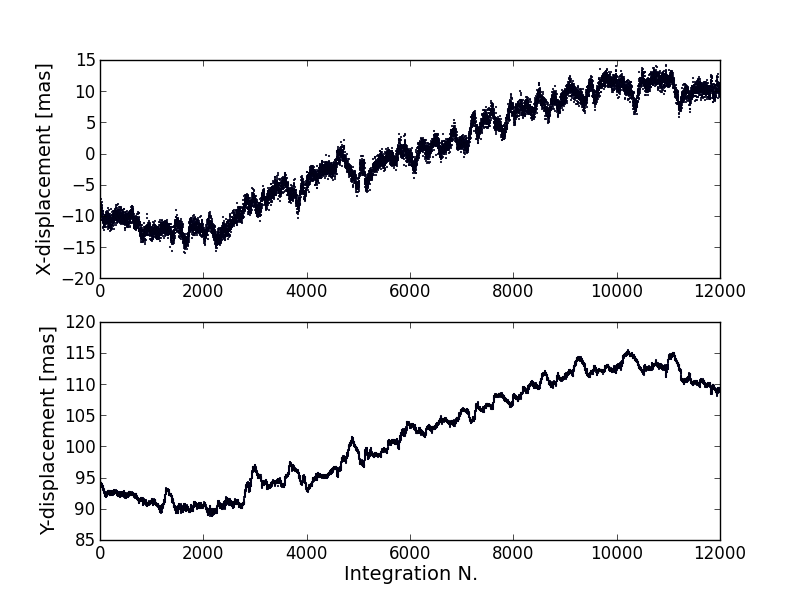}

\caption{\label{fig:shifts}Jitter and drift of the source
  that was used to mimic the observation with NIRSpec of a bright star
  for a three hour long exposure during the cryogenic ground test campaign.}
\end{figure}

The continuum light-source consisted of a tungsten filament with a PSF
similar to that of JWST. Due to the test environment, the source
underwent spatial jitter and a drift of more than 30 mas over the
3-hour exposure. We traced the source displacement over all the
integrations measuring the baricenter position of the spectral trace,
for the source movements in cross-dispersion direction
(that is, the y-direction), and the center of a prominent carbon-dioxide
absorption feature at 4.2 $\mu$m, for displacements in dispersion
direction (that is, the x-direction). These measurements are shown in
Fig.\,\ref{fig:shifts}, and we note that the amplitude of the drift is more
than four times the pointing stability requirement for JWST, which has
to be better than 7 mas (1$\sigma$) over 10,000 s.

\begin{figure}
\centering
\includegraphics[width=9 cm]{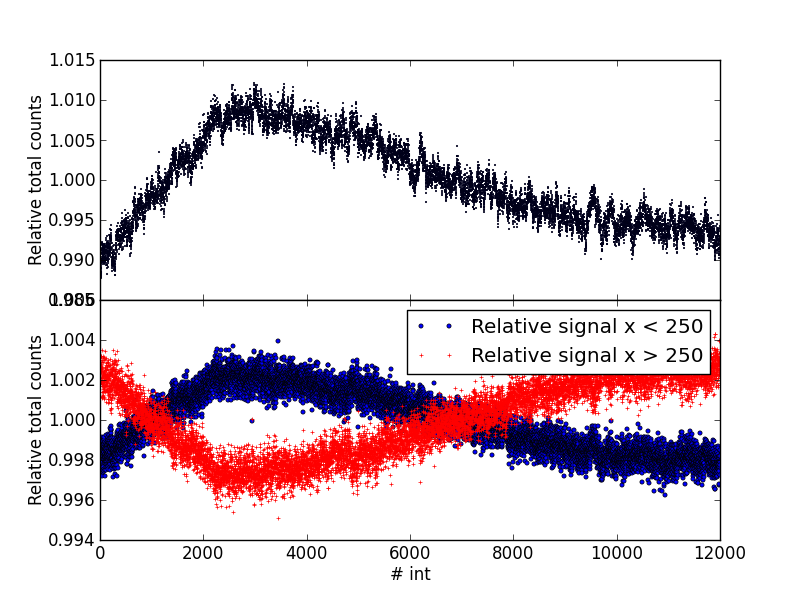}

\caption{\label{fig:lamp_flux} Relative white-light curve for each integration. Obtained by summing up the signal of pixels in the trace and normalizing by the average value over all integrations. Top panel: for all pixels in the trace. Bottom Panel: for pixels in the trace with coordinate in the dispersion direction, $x < 250$ (blue dots) and $x > 250$ (red crosses), normalized by the white light curve, so that the values trace the source flux variation in the blue and red part of the spectrum relative to the total flux.}
\end{figure}

The first step of NIRSpec data analysis consists in producing
count rate images from the up-the-ramp individual integrations, so
12,000 count rate images were produced from the raw data by our
standard ramp-to-slopes pipeline (see Section~\ref{sec:pipe}). We do not expect the source
luminosity to be stable over the exposure time-scale and it is clear
that the standard deviation ($1\sigma$) over the 12,000 integrations of the illuminated pixels in the count rate images is dominated by the overall total counts
variations as traced by the white light curve, that is, the total
signal in the detector subarray for each integration, as shown in the
top-panel of Fig.\,\ref{fig:lamp_flux} (normalized to the signal
average over the all exposure). 

Therefore, after having produced the count rate image for each
integration, assigned a wavelength to each pixel in the image, and
spatially collapsed the 2D data set, the spectrum so derived for each
integration was normalized by the relative white-light curve value (as
given in Fig.\,\ref{fig:lamp_flux}). After this correction, it is
apparent that the $1\sigma$ of each pixel over the 12,000 spectra is
dominated by the contribution of the correlated noise from the
detector read-out electronics. This noise component manifests itself in
low level cross-dispersion stripes and bands visible in the count rate
images. The presence of these fluctuations in Teledyne HAWAII
detectors is well known and the special read-out mode NRSIRS2, in
conjunction with a Wiener filtering algorithm named IRS$^2$ (to be
applied by the ramp-to-slope processing pipeline), has been developed
to minimize the impact of this $1/f$-noise source in full-frame mode \citep[][and also Paper I]{maf+2010,raf+2012}.

Reference pixels are not always available when using detector subarrays depending on the subarray size and its location, however in the case of the observation of a point source with a subarray of size 32 pixels in spatial direction (such as \verb+SUB512+ or \verb+SUB2048+), there is a significant number of pixels in each column that are not illuminated and can be used to track the $1/f$-noise components, typically more than ten. To filter out this noise components from the count rate images we used a basic algorithm based on computing the median count rate of the dark pixels in a column (i.e., all pixels with projected spatial coordinate $y >|0.5| S_y$, where $S_y$ is the slit size in spatial direction), and subtracting this value from the count rate of all the pixels in that column. The operation is repeated for each column on every count rate images in the 12,000 sequence.

\begin{figure}
\centering
\includegraphics[width=9 cm]{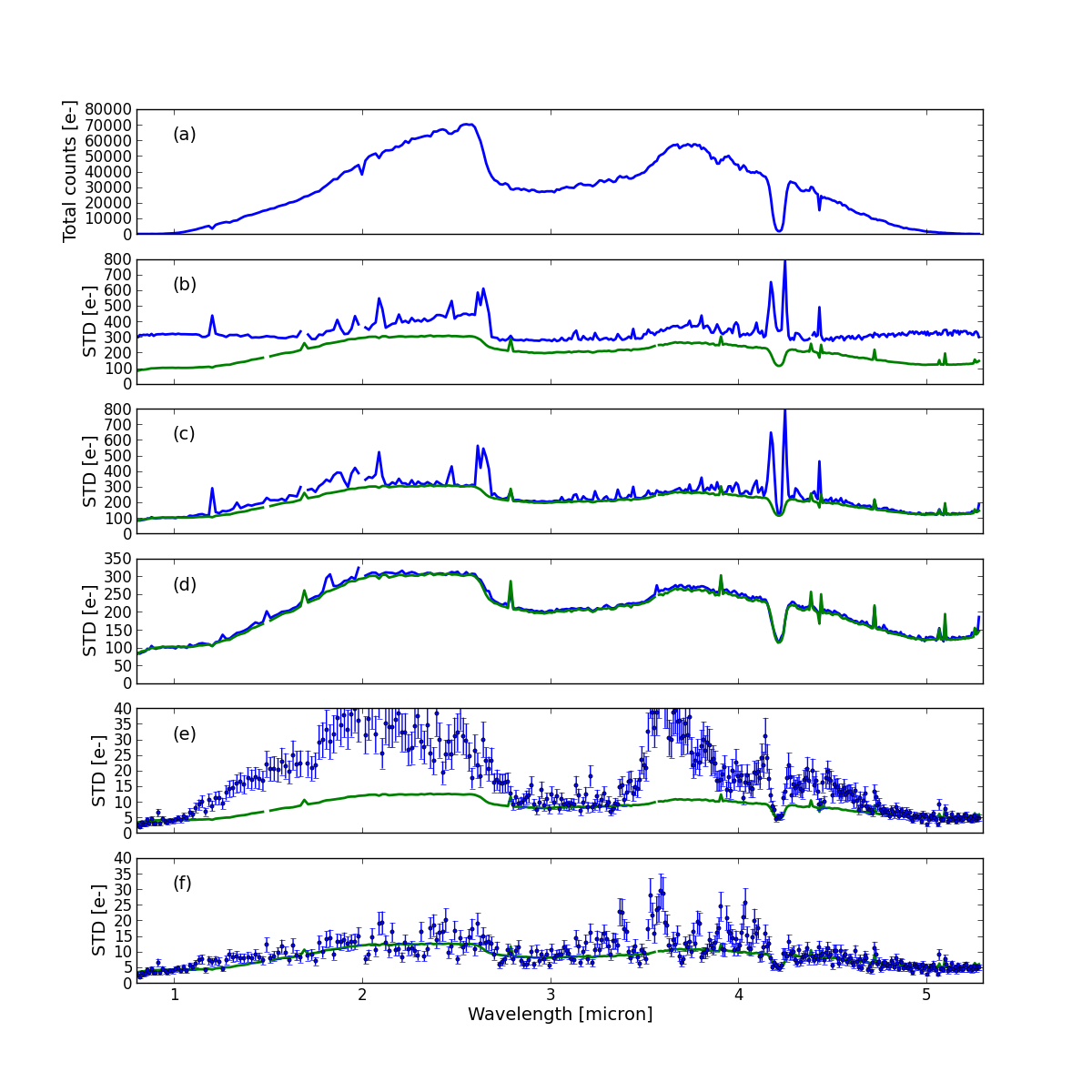}

\caption{\label{fig:multi} Spatially collapsed spectrum of the bright point source observed with NIRSpec during ISIM CV3. The signal is shown in panel {\em a}, all other panels show the $1\sigma$ deviation of each pixel in the collapsed trace. Shown in green is the expected level of $1\sigma$ from the statistical noise sources of an ideal instrument and in blue the measured $1\sigma$ over the last 9,000 integrations of the three hour long exposure. Panel {\em b} shows the $1\sigma$ of the raw data, white-light curve normalized; panel {\em c} gives $1\sigma$ like {\em b}, but after image destriping
  ($1/f$-noise removal using dark pixels) and flat-fielding; panel {\em d} is as {\em c}, but after having decorrelated each pixel signal by the source spatial shifts in $x$
  and $y$ direction; panel {\em e} is as {\em d}, but here the $1\sigma$ is computed over 15 (temporal bins) of 600 integrations each, allowing us to reach a higher sensitivity. Error bars on the $1\sigma$ are also given. Panel {\em f} is as {\em e}, but rather than normalizing by the white light-curve, here the data were corrected by a five-color light
  curve (see text).}
\end{figure}

Despite its simplicity, this ``destriping'' algorithm is very
effective in suppressing the $1/f$-noise. Our ramp-to-slope pipeline
provides, together with the count rate image of each integration, the
statistical variance for each pixel computed from the pixel count rate
and the expected variance of the read noise\footnote{as measured for each pixel from the correlated double sampe (CDS) noise in dark exposures} using the expression given Eq.\,1 of
\cite{rff07,rff10err}. Starting from this value we can derive the
expected (due to statistical fluctuations only) $1\sigma$ of our signal, in
each pixel of the collapsed subarray for a given integration time.  As
illustrated in Fig.\,\ref{fig:multi}, before destriping, the measured
$1\sigma$ of the total electron count in each integration is 3-4 times the
expected value, depending on the incident signal (panel {\em b}),
while after filtering out the $1/f$-noise component, the observed $1\sigma$
closely matches the ideal level in the absence of signal (see panel
{\em c} for $\lambda < ~1.0~\mu{\rm m}$ and $\lambda > 4.8~\mu{\rm
  m}$). 
  
Nevertheless, a large excess of noise remains in the presence of
signal and in particular around the steepest spectral gradients.
This excess is due to the interplay between the under-sampled
PSF, whose FWHM is comparable to the pixel size, and the spatial
movements of the source. To correct for this effect, after
flat-fielding, we performed a bilinear fit of each pixel signal
(operating on the spectral 2D trace) as a function of the source
position in x and y in each integration (as plotted in
Fig.\,\ref{fig:shifts}) and subtracted the fitted value from the pixel
value. As shown in Panel {\em d} of Fig.\,\ref{fig:multi}, this (first
order) correction brings the measured $1\sigma$ over 9,000 integrations in
line with the expected value, that is, between 200 and 300 $e^{-}$ for signal
> 30,000 $e^{-}$.  The first 3,000 integrations of the exposure were
not included in this analysis because the test source had just been
switched on and exhibited a more prominent (wavelength dependent)
drift than the white-light-curve normalization is able to correct
for (see Fig.\,\ref{fig:lamp_flux} and the discussion below).

In order to reach higher sensitivity we binned the 9,000 integrations
into 15 bins of 600 each, corresponding to time periods of 4.5 minutes. For
this integration time the expected $1\sigma$ of a signal between
30,000$-$70,000 $e^{-}$ is 8$-$12 $e^-$, corresponding to a
noise-to-signal ratio (N/S) of less than 300 ppm per pixel, however,
as shown in Panel {\em e} of Fig.\,\ref{fig:multi}, the measured $1\sigma$
are in some wavelength ranges in excess of two times these values. While
we cannot exclude that other instrumental effects are causing (part
of) the excess noise at this level of sensitivity, the $1\sigma$ measured
here is very likely dominated by the flux instability of the light
source used in the test.

We evaluated the relative variations of the blue and red parts of the
spectrum by taking the total signal in the blue half of the subarray
($x <250$) and in the red-part ($x> 250$), as shown in Fig.\,\ref{fig:lamp_flux} (bottom panel). From integration 3,000 onward, the relative amplitude of the blue and red light curve are monotonically varying by $\sim$ 0.6 \% relative to the total flux
drift. 

There are various reasons to be confident that here we are
genuinely looking at the lamp flux variations as opposed to trends in
the detector response. After filtering out the $1/f$-component from
the detector noise, the $1\sigma$ of the dark pixels is fully consistent
with the expected value as based on the read noise and dark current,
excluding intrinsic variations of the additive type of the
detector response. Beside the PRISM data discussed here, we also acquired a
3-hour exposure with grating G235H dispersing the light from the same
source onto both NIRSpec detectors (operated in window-mode with window
of size 2048$\times$32). Cross-correlating the white light curves of
the data from NRS1 to NRS2, we find highly correlated data with a
time lag that corresponds to the difference in exposure start times of the two detectors. This excludes independent response fluctuations of a
multiplicative type (that is, proportional to the flux) in each 
detector or response variations synchronous in both detectors such as
the type that could possibly be driven by drifts in operating
temperature.

To test the assumption that the excess noise seen here is due to the color changes of the lamp, we subdivided the collapsed spectral data into five bins of 40 pixels each and computed the total signal in each bin for each integration (thereby producing five light
curves of 9,000 points). Before computing the (temporally binned) $1\sigma$ for each pixel, the time-series for each pixel was normalized by a light curve as derived by interpolating between the two spectrally adjacent light curves, integration by integration. The values of the $1\sigma$ for each pixel derived in this way are shown in the bottom panel of Fig.\,\ref{fig:multi}, with the measured $1\sigma$ now more in line with the expected values.

The five-color-light curve normalization allows us to reach the expected
sensitivity level of 200$-$300 ppm, for total counts of
70,000$-$30-000, although higher residual noise is left in the
wavelength interval 3.4$-$4 $\mu$m, where the five-color correction does
not perform as well. It is clear from this analysis that the
performance of the source in terms of flux stability limits our
ability to probe possible sources of systematic noise in NIRSpec using
this data set. Nevertheless, the fact that the residual noise above
the floor level can be reduced here by a simple tracking of the source
flux in five wavelength bins, give us confidence in the stability of
NIRSpec over the three hour time-scale typical of a transit
observation and the ability of the instrument to reach the noise-floor
of 200 ppm in less than five minutes of integration. The median of the difference between
ideal and measured $1\sigma$ of each pixel over $\sim$ 200 pixels with
signal greater than 30,000~$e^-$, is $\sim$ 2 $e^-$, indicating that
NIRSpec performance in terms of noise level is within 20\% of the
ideal value, for integration times of five minutes.

\end{appendix}

\end{document}